\documentclass[aip,jcp,amsmath,amssymb,preprint]{revtex4-1}
\usepackage{bm}
\usepackage{mathrsfs}

\usepackage{latexsym,amsmath, amscd, amssymb} 
\usepackage{color,calc,array,epic,subfigure}
\usepackage[dvips]{epsfig,graphicx}
\usepackage{setspace}
\usepackage[dvips]{graphicx}
\usepackage[normalem]{ulem}
\graphicspath{{/Users/bl/Dropbox/isok-submit/figs/}}
\newcommand{\rx}{{\rm x}}
\newcommand{\ur}{{\bf r}}

\newcommand{\dt}{\Delta t}
\newcommand{\dti}{\delta t}

\def\dd{{d}}

\begin{document}

\title{Stochastic, resonance-free multiple time-step algorithm for molecular dynamics
with very large time steps}

\author{Ben Leimkuhler}
\affiliation{The Maxwell Institute and School of Mathematics, University
of Edinburgh EH9 3JZ, United Kingdom}
\author{Daniel T. Margul}
\affiliation{Department of Chemistry, New York University, New York, NY 10003, USA}
\author{Mark E. Tuckerman}
\affiliation{Department of Chemistry and Courant Institute of Mathematical Sciences, New York University, New York, NY 10003, USA}

\email{mark.tuckerman@nyu.edu}

\date{\today}

\begin{abstract}
Molecular dynamics is one of the most commonly used approaches for studying
the dynamics and statistical distributions of many
physical, chemical, and biological systems using atomistic or coarse-grained models.  
It is often the case, however, that the interparticle forces
drive motion on many time scales, and 
the efficiency of a calculation is limited by
the choice of time step, which must be sufficiently small
that the fastest force components are accurately integrated.  Multiple
time-stepping algorithms partially alleviate this inefficiency by
assigning to each time scale an appropriately chosen step-size.  However,
such approaches are limited by resonance phenomena, wherein
motion on the fastest time scales limits the step sizes
associated with slower time scales.  In atomistic models of biomolecular
systems, for example, resonances limit the largest time step to 
around 5-6 fs. Stochastic processes promote mixing and ergodicity in dynamical systems and reduce the impact of resonant modes.  In this paper, we introduce a set of stochastic isokinetic 
equations of motion that are shown to be rigorously ergodic and that can be integrated using a
multiple time-stepping algorithm which is easily implemented 
in existing molecular dynamics codes.
The technique is applied to a simple, illustrative problem and then to a more realistic
system, namely, a flexible
water model.  Using this approach outer time steps as large as 100 fs are 
shown to be possible.
\end{abstract}

\maketitle

\section{Introduction}

Sampling the conformational equilibria of complex systems remains a major
challenge in molecular simulation.  If achieved, problems such as
biomolecular structure prediction, exploration of polymorphism in 
molecular crystals, and determination of equilibria in glassy systems,
to name just a few, would be significantly impacted.  Numerous approaches
have been developed to accelerate sampling of the Boltzmann distribution,
many of which employ molecular dynamics (MD) as the basic engine for
driving the exploration of configuration 
space~\cite{Huber94,Grubmuller95,Darve01,FREE_ADB_MET1,FREE_ADB_MET2,FREE_ADB_MET3,FREE_ACTION4,Henin04,SIAM_07,abf2,TAMD,D-AFED,Dickson10,Yu11,UFED}.  
However, methods
such as these do not address the underlying inefficiency associated with 
the inherent broad range of time scales present in the interparticle 
forces.  Multiple time step (MTS) 
integrators~\cite{Tuckerman92,Zhou_1995,Zhou_1996,Martyna96,Zhou_1997,Omelyan09,Morrone_2010,Morrone_2011} 
were designed to address this problem, and they are capable of improving the efficiency
of MD-based approaches even further.  However, because MTS methods 
are essentially perturbative in nature, their efficiency is strongly
limited by resonance phenomena~\cite{Schlick_98,Ma_03}, which limit the largest time step to
$\Delta t \lesssim 5$ fs for many types of problem, including biomolecular
systems treated with an atomistic model.  
The resonance problem thus poses a major bottleneck for MTS schemes.

Some time ago, Minary {\it et al.} introduced an MTS algorithm which eliminates (or at least controls)
resonance phenomena~\cite{Minary_PRL}.  The algorithm is an
extended phase-space method based on the 
Nos\'e-Hoover chain approach (NHC)~\cite{Martyna92} combined with a set of
isokinetic constraints~\cite{Evans_Morriss,Zhang_97,TUCK_ISOK1,Tuckerman_SM} 
that couple the physical degrees of freedom to
the NHC thermostats. The isokinetic constraints restrict the energy
that can build up in any one mode of motion, thereby moderating the effect of the resonance.  
The scheme is termed ``Iso-NHC-RESPA'', as it starts from the original 
reference system propagator algorithm (RESPA) introduced
by Tuckerman {\it et al.}~\cite{Tuckerman92} and builds in
the reversible isokinetic integration algorithms introduced
by Zhang~\cite{Zhang_97} and by Tuckerman {\it et al.}~\cite{TUCK_ISOK1}
and the reversible NHC integrators of Martyna {\it et al.}~\cite{Martyna96}.
Using the Iso-NHC-RESPA
approach, MD simulations for conformational sampling can be performed
with outer time steps as large as 100 fs, and in rigid body MD, even
larger outer time steps are possible~\cite{Omelyan_2011}.  It is important to note, however,
that the dynamics are not preserved in the Iso-NHC-RESPA scheme due 
to the strongly non-Hamiltonian nature of the underlying equations
of motion~\cite{MET_Nonham1,MET_Nonham2} resulting from a large number
of isokinetic constraints.
More recently, Morrone {\it et al.} showed that
coloured noise thermostats~\cite{Morrone_2011} can be used to devise resonant-free
MTS algorithms that are able to preserve the dynamics, although
the gains in time step are more modest.  

In Ref.~\onlinecite{Minary_PRL}, it was suggested that each physical degree of freedom
be coupled to its own NHC thermostat through an isokinetic constraint
in a ``massive'' thermostatting approach.    The massive thermostats, as proposed, have two notable
limitations
\begin{itemize}
\item Since the thermostat is applied to each degree of freedom, the integration of the many NHC thermostats carries an overhead.
\item Although in many practical applications NHC methods appear to be ergodic, no proof
exists of this fact, hence also the ergodicity of the fully deterministic massive, isokinetic NHC method must be regarded as not established from a mathematical perspective.
\end{itemize}
Both of these issues can be addressed by introducing a stochastic modification of the equations in 
place of the thermostatting chain \cite{SaDeCh07}.   This has been referrred to as the 
{\em Nos\'{e}-Hoover-Langevin} (NHL) method in Ref. \onlinecite{LeNoTh09}, where it was also proved
to be ergodic when applied to linear systems.

In this paper, we show that the Iso-NHC-RESPA scheme can be reformulated
as a resonant-free MTS algorithm replacing the thermostat chains with 
the stochastic NHL scheme, thereby allowing an increase in the efficiency
of the approach without a reduction in the outer time step.    We refer to this method as  Stochastic-Iso-NH-RESPA or SIN(R)
for short.  The ``R'' for ``RESPA'' is included parenthetically as a reminder that
the SIN(R) scheme can be used purely as a thermostatting method without inclusion of 
RESPA multiple time stepping.  However, the real advantage of the method will be demonstrated
when it is employed as an MTS approach.

The paper is organized as follows.   In Sec.~\ref{sec:eoms}, we give the equations of motion.  We include a formulation incorporating a chain of auxiliary thermostatting variables as well as a stochastic perturbation.   Sec.~\ref{sec:analysis} contains an analysis of the properties of the method, including (a) the study of its invariant distribution based on the non-Hamiltonian statistical mechanical formalism (refs.~\onlinecite{MET_Nonham1,MET_Nonham2}), and (b) a discussion of the ergodicity of method (for a harmonic model problem) using the H\"ormander condition. 
In Sec.~\ref{sec:integrator}, we provide the details
of a recommended numerical integrator for the equations of motion
and discuss its advantages over other possible schemes.  
In Sec.~\ref{sec:applications}, we provide a number of numerical
examples and demonstrate the performance of the method.  In this
section, we also examine the errors associated with different
choices of the parameters.  Conclusions are given in 
Sec.~\ref{sec:concl}. 

\section{Equations of motion}
\label{sec:eoms}

Consider a system of $N$ particles in $d$ dimensions with 
coordinates $q_1,...,q_{dN}\equiv q$, momenta $p_1,...,p_{dN} \equiv p$, and masses
$m_1,...,m_N$.  The interparticle forces present between
the particles are denoted $F_1(q),...,F_{dN}(q)$, and these
are assumed to give rise to motion on a wide variety of time
scales ranging from fast bond vibrations to slowly varying
long-range electrostatic and van der Waals components
and are derived from an $N$-particle potential $U(q_1,...,q_{dN})\equiv U(q)$.
Let $\alpha=1,...,dN$, let $v_{\alpha} = \dot{q}_{\alpha}$
be the velocity associated with $q_{\alpha}$, and let $m_{\alpha}$ be
associated mass.
For each coordinate in the system, we introduce the following set of stochastic
equations of motion:
\begin{eqnarray}
dq_{\alpha} & = & v_{\alpha}dt
\nonumber \\
dv_{\alpha} & = & \left[{F_{\alpha}(q) \over m_{\alpha}} - \lambda_{\alpha} v_{\alpha}\right]dt
\nonumber \\
dv_{1,\alpha} & = & -\lambda_{\alpha} v_{1,\alpha} dt - v_{2,\alpha} v_{1,\alpha} dt
\nonumber \\
dv_{2,\alpha} & = & {G(v_{1,\alpha}) \over Q_2}dt -\gamma v_{2,\alpha}dt + \sigma d{\rm W_{\alpha}}
\label{eq:eoms4}
\end{eqnarray}
where $v_{1,\alpha}$ and $v_{2,\alpha}$ are auxiliary thermostat variables,
$\sigma = \sqrt{2\gamma/\beta Q_2}$, $\gamma$ is 
the friction constant,
$\beta = 1/k_{\rm B}T$, $T$ is the temperature of the system,
$k_{\rm B}$ is Boltzmann's constant, 
\begin{equation}
G(v) = Q_1 v^2 - \beta^{-1}
\end{equation}
$Q_1$ and $Q_2$ are
thermostat coupling parameters,
and $\lambda_{\alpha}$ is a Lagrange multiplier
for enforcing an isokinetic constraint between the
physical velocity $v_{\alpha}$ and the thermostat velocity $v_{1,\alpha}$
{\it on each degree of freedom}:
\begin{equation}
mv_{\alpha}^2 + {1 \over 2}Q_1 v_{1,\alpha}^2 = \beta^{-1}
\label{eq:isokin}
\end{equation}
As will be seen in Sec.~\ref{sec:analysis}, this form of the
isokinetic constraint ensures that a proper canonical distribution
in the coordinates is generated.  Note, however, that
the individual values of $mv_{\alpha}^2$ and $Q_1v_{1,\alpha}^2$ are
not what one would expect from a canonical distribution in the velocities
but rather correspond to a particular type of microcanonical distribution
in each subset $\{v_{\alpha},v_{1,\alpha}\}$ of particle velocities~\cite{Tuckerman_SM}.
An explicit expression for the Lagrange multiplier can be obtained by
differentiating Eq.\ (\ref{eq:isokin}) with respect to time, which yields
\begin{equation}
2mv_{\alpha}\dot{v}_{\alpha} + Q_1v_{1,\alpha}\dot{v}_{1,\alpha} = 0,
\end{equation}
substituting
the equations of motion in for $\dot{v}_{\alpha}$ and $\dot{v}_{1,\alpha}$ to give
\begin{equation}
2mv_{\alpha}\left[{F_{\alpha}(q) \over m_{\alpha}} - \lambda_{\alpha} v_{\alpha}\right]
+ Q_1 v_{1,\alpha}\left[-\lambda_{\alpha}v_{1,\alpha} - v_{2,\alpha}v_{1,\alpha}\right] = 0,
\end{equation}
and then solving for $\lambda_{\alpha}$.   The result is
\begin{equation}
\lambda_{\alpha} = {v_{\alpha}F_{\alpha}(q) - {1 \over 2}Q_1v_{1,\alpha}^2v_{2,\alpha} \over
mv_{\alpha}^2 + {1 \over 2}Q_1 v_{1,\alpha}^2}
\label{eq:lambda1}
\end{equation}
 
 The symbol $d {\rm W_{\alpha}}$ appearing in Eq. (\ref{eq:eoms4}) stands for the infinitesimal increment of a Wiener process ${\rm W_{\alpha}}(t)$, that is, a ``white noise" stochastic process,  continuous in $t$, and such that ${\rm W_{\alpha}}(t)-{\rm W_{\alpha}}(s)$ is normally distributed with mean zero and variance $|t-s|$, whereas ${\rm W}_{i}$ and ${\rm W}_{j}$ are independent processes if $i\neq j$.  The equations are thus taken to be a system of stochastic differential equations in the Ito sense.
The coupling between $v_{1,\alpha}$ and $v_{2,\alpha}$ is of the Nos\'e-Hoover type
with a Langevin-like driving of the variable $v_{2,\alpha}$ 
(similar to the Nos\'{e}-Hoover-Langevin method \cite{SaDeCh07,LeNoTh09}).
If Eq.\ (\ref{eq:lambda1}) is substituted into Eqs.\ (\ref{eq:eoms4}), 
the resulting equations of motion become
\begin{eqnarray}
dq_{\alpha} & = & v_{\alpha}dt
\nonumber \\
\nonumber \\
dv_{\alpha} & = & \left[{F_{\alpha}(q) \over m_{\alpha}} - 
{v_{\alpha}F_{\alpha}(q) - {1 \over 2}Q_1v_{1,\alpha}^2v_{2,\alpha} \over
mv_{\alpha}^2 + {1 \over 2}Q_1 v_{1,\alpha}^2}v_{\alpha}\right]dt
\nonumber \\
\nonumber \\
dv_{1,\alpha} & = & -{v_{\alpha}F_{\alpha}(q) - {1 \over 2}Q_1v_{1,\alpha}^2v_{2,\alpha} \over
mv_{\alpha}^2 + {1 \over 2}Q_1 v_{1,\alpha}^2}
v_{1,\alpha} dt - v_{2,\alpha} v_{1,\alpha} dt
\nonumber \\
\nonumber \\
dv_{2,\alpha} & = & {G(v_{1,\alpha}) \over Q_2}dt -\gamma v_{2,\alpha}dt + \sigma d{\rm W_{\alpha}}
\label{eq:eoms4p}
\end{eqnarray}
From Eqs.\ (\ref{eq:eoms4p}), an important fact about the equations of motion can be
derived.  In particular, if we use the fact that 
$Q_vv_{1,\alpha}^2/2 = \beta^{-1} - mv_{\alpha}^2$, the equation of motion
for $v_{1,\alpha}$ can be expressed as
\begin{equation}
dv_{1,\alpha} = -\left[{v_{\alpha}F_{\alpha}(q) - \left(\beta^{-1} - mv_{\alpha}^2\right) - \beta^{-1}v_{2,\alpha}
\over \beta^{-1}}\right]v_{1,\alpha}dt
\end{equation}
which can also be written as
\begin{equation}
d\ln |v_{1,\alpha}| = -\left[{v_{\alpha}F_{\alpha}(q) - \left(\beta^{-1} - mv_{\alpha}^2\right) - \beta^{-1}v_{2,\alpha}
\over \beta^{-1}}\right]dt
\label{eq:v1_domain}
\end{equation}
The consequence of Eq.\ (\ref{eq:v1_domain}) is that $v_{1,\alpha}$ can never change sign.
Hence, its domain is either $v_{1,\alpha} \geq 0$ or $v_{1,\alpha} \leq 0$.

As was done in Ref.~\onlinecite{Minary_PRL}, it is possible to generalize Eqs.\ (\ref{eq:eoms4})
to incoprorate a set of $2L$ thermostat variables, which slightly
modifies the form of the equations of motion to read:
\begin{eqnarray}
dq_{\alpha} & = & v_{\alpha}dt 
\nonumber \\
dv_{\alpha} & = & \left[{F_{\alpha}(q) \over m_{\alpha}} - \lambda_{\alpha} v_{\alpha}\right]dt
\nonumber \\
dv_{1,\alpha}^{(k)} & = & -\lambda_{\alpha} v_{1,\alpha}^{(k)} dt - v_{2,\alpha}^{(k)} v_{1,\alpha}^{(k)} dt
\nonumber \\
dv_{2,\alpha}^{(k)} & = & {G(v_{1,\alpha}^{(k)}) \over Q_2}dt -\gamma v_{2,\alpha}^{(k)}dt + \sigma dW_{\alpha}^{(k)}
\label{eq:eomsk}
\end{eqnarray}
with the isokinetic constraint taking the form
\begin{equation}
mv_{\alpha}^2 + {L \over L+1}\sum_{k=1}^L Q_1 \left(v_{1,\alpha}^{(k)}\right)^2 = L\beta^{-1}
\label{eq:isokink}
\end{equation}
Again, the form of this constraint is chosen to give a proper canonical distribution
in the coordinates as we will show in Sec.~\ref{sec:analysis}~\cite{Minary_PRL}.
There it will be shown that the distribution generated is a product
of microcanonical distributions in the subsets $\{v_{\alpha},v_{1,\alpha}^{(k)}\}$ 
for each degree of freedom.  Following the procedure outlined previously of 
differentiating the constraint once with respect to time and substituting 
the equations of motion in for the time derivatives, we find that
the expression for the Lagrange multiplier generalizes to
\begin{equation}
\lambda_{\alpha} = {v_{\alpha}F_{\alpha}(q) - {L \over L+1}\sum_{k=1}^L Q_1\left(v_{1,\alpha}^{(k)}\right)^2v_{2,\alpha}^{(k)} \over
mv_{\alpha}^2 + {L \over L+1}\sum_{k=1}^L Q_1 \left(v_{1,\alpha}^{(k)}\right)^2}
\label{eq:lambdak}
\end{equation}
Note that Eq.\ (\ref{eq:isokink}) reduces to Eq.\ (\ref{eq:isokin}) when $L=1$.  Moreover, an argument
similar to that lead to Eq.\ (\ref{eq:v1_domain}) can also be made for this more general case,
and it leads to the same restriction on the domain of $v_{1,\alpha}^{(k)}$, {\it i.e.,}
$v_{1,\alpha}^{(k)} \geq 0$ or $v_{1,\alpha}^{(k)} \leq 0$. 
Concerning the thermostat coupling
parameters $Q_1$ and $Q_2$, we recommend that these be choen in a manner similar to the corresponding parameters
of Nos\'e-Hoover chains~\cite{Martyna92,Martyna96}, {\it i.e.,} 
$Q_1 = Q_2 = k_{\rm B}T\tau^2$, where $\tau$ is a time scale of relevance in the system.
However, as we will see in Sec.~\ref{sec:applications},
the ability of Eqs.\ (\ref{eq:eoms4}) or (\ref{eq:eomsk}) to produce a canonical configurational distribution
is not particularly sensitive to this choice.

\section{Analysis of the extended isokinetic system}
\label{sec:analysis}
In this section, we discuss Eqns.\ (\ref{eq:isokink}), showing  that they generate
a canonical distribution in the configuration space of the
$dN$ coordinates $q_1,...,q_{dN}$.   The analysis proceeds in two stages: first, the demonstration that the method preserves the isokinetic partition function, and second that it is ergodic.  The proof of ergodicity is only supplied here for $L=1$.  Extending the proof of ergodicity to arbitrary $L$ would be possible but technically involved, and the property is unlikely to fail as the $L>1$ case only increases the interaction of the variables compared to $L=1$.

\subsection{Preservation of the isokinetic distribution}
Here we employ a procedure derived by
Tuckerman {\it et al.}~\cite{MET_Nonham1,MET_Nonham2}.  In this scheme, 
if a set of dynamical equations $\dot{\rx} = \xi(\rx)$,
where $\rx$ is the full phase-space vector (including any
extended phase-sapce variables) and $\xi(\rx)$ is 
a vector field, has $N_c$ conservation
laws of the form $\Lambda_l(\rx) = C_l$, $l=1,...,N_c$, 
then, assuming the motion is ergodic, the equations of motion
generate a generalized microcanonical distribution for which
the corresponding partition function takes the form
\begin{equation}
\Omega(C_1,...,C_{N_c}) = \int d\rx \sqrt{g(\rx)}
\prod_{l=1}^{N_c}\delta\left(\Lambda_l(\rx) - C_l\right)
\label{eq:pfn}
\end{equation}
Here, $\sqrt{g(\rx)}d\rx$ is a conserved volume element, and 
$\sqrt{g(\rx)}$ is a metric factor determined by the
compressibility of the equations of motion, which is given by
\begin{equation}
\kappa = \nabla_{\rx}\cdot \xi(\rx)
\label{eq:compress}
\end{equation}

An explicit  form for the equations of motion can be obtained
by substituting Eq.\ (\ref{eq:lambdak}) into
Eqs.\ (\ref{eq:eomsk}), which yields
\begin{eqnarray}
dq_{\alpha} & = & v_{\alpha} dt
\nonumber \\
\nonumber \\
dv_{\alpha} & = & {F_{\alpha}(q) \over m_{\alpha}}dt - 
\left[{v_{\alpha}F_{\alpha}(q) - {L \over L+1}\sum_{k=1}^L Q_1\left(v_{1,\alpha}^{(k)}\right)^2v_{2,\alpha}^{(k)} \over
mv_{\alpha}^2 + {L \over L+1}\sum_{k=1}^L Q_1 \left(v_{1,\alpha}^{(k)}\right)^2}\right]v_{\alpha}dt
\nonumber \\
\nonumber \\
dv_{1,\alpha}^{(k)} & = & -\left[{v_{\alpha}F_{\alpha}(q) - {L \over L+1}\sum_{j=1}^L Q_1\left(v_{1,\alpha}^{(j)}\right)^2v_{2,\alpha}^{(j)} \over
mv_{\alpha}^2 + {L \over L+1}\sum_{j=1}^L Q_1 \left(v_{1,\alpha}^{(j)}\right)^2}\right]
v_{1,\alpha}^{(k)} dt - v_{2,\alpha}^{(k)} v_{1,\alpha}^{(k)} dt
\nonumber \\
\nonumber \\
dv_{2,\alpha}^{(k)} & = & {G(v_{1,\alpha}^{(k)}) \over Q_2}dt -\gamma v_{2,\alpha}^{(k)}dt + \sigma dW^{(k)}
\label{eq:econsk_fin}
\end{eqnarray}

The next step is the calculation of the phase-space compressibility.  
For this analysis, we first remove the friction and
random force terms from the $v_{2,\alpha}^{(k)}$ equation (we return to this point below).

%
%
The
compressibility is given by
\begin{equation}
\kappa  = \sum_{\alpha=1}^{3N}
\left[{\partial \dot{q}_{\alpha}
\over \partial q_{\alpha}} + 
{\partial \dot{v}_{\alpha}
\over \partial v_{\alpha}} + 
\sum_{k=1}^L
\left({\partial \dot{v}_{1,\alpha}^{(k)}
\over \partial v_{1,\alpha}^{(k)}}
+ {\partial \dot{v}_{2,\alpha}^{(k)}
\over \partial v_{2,\alpha}^{(k)}}\right)\right]
\label{eq:kappa_def}
\end{equation}
The equations of motion we will use in the calculation 
of the compressibility are simply Eqs.\ (\ref{eq:econsk_fin})
with the friction and random force terms removed, which are
\begin{eqnarray}
\dot{q}_{\alpha} & = & v_{\alpha}
\nonumber \\
\nonumber \\
\dot{v}_{\alpha} & = & {F_{\alpha}(q) \over m_{\alpha}} - 
\left[{v_{\alpha}F_{\alpha}(q) - {L \over L+1}\sum_{k=1}^L Q_1 \left(v_{1,\alpha}^{(k)}\right)^2 v_{2,\alpha}^{(k)} \over
mv_{\alpha}^2 + {L \over L+1}\sum_{k=1}^L Q_1 \left(v_{1,\alpha}^{(k)}\right)^2}\right]v_{\alpha}
\nonumber \\
\nonumber \\
\dot{v}_{1,\alpha}^{(k)} & = & -\left[{v_{\alpha}F_{\alpha}(q) - {L \over L+1}\sum_{j=1}^L Q_1\left(v_{1,\alpha}^{(j)}\right)^2v_{2,\alpha}^{(j)} \over
mv_{\alpha}^2 + {L \over L+1}\sum_{j=1}^L Q_1 \left(v_{1,\alpha}^{(j)}\right)^2}\right]
v_{1,\alpha}^{(k)} - v_{2,\alpha}^{(k)} v_{1,\alpha}^{(k)}
\nonumber \\
\nonumber \\
\dot{v}_{2,\alpha}^{(k)} & = & {G(v_{1,\alpha}^{(k)}) \over Q_2}. \nonumber
\label{eq:eoms_an}
\end{eqnarray}
These equations possess $dN$ conservation laws of the form
\begin{equation}
\Lambda_{\alpha} = mv_{\alpha}^2 + {L \over L+1}\sum_{k=1}^L Q_{1}\left(v_{1,\alpha}^{(k)}\right)^2 = L\beta^{-1}.
\end{equation}
For the purposes of this analysis, we take $m_{\alpha}=1$ and 
$Q_{1} = Q_{2} = 1$.  
After some straightforward but tedious algebra, the compressibility 
can be shown to be
\begin{equation}
\kappa = \sum_{\alpha=1}^{3N}
\left[{L \over \Lambda_{\alpha}}\left(-F_{\alpha}(q)v_{\alpha}
+ \sum_{k=1}^L v_{2,\alpha}^{(k)}\left(v_{1,\alpha}^{(k)}\right)^2\right)
- \sum_{k=1}^L v_{2,\alpha}^{(k)}\right].
\end{equation}
Using the fact that $\Lambda_{\alpha} = L/\beta$, the compressibility can be expressed
as a total time derivative
\begin{equation}
\kappa = \beta \left [ {dU \over dt} + {d \over dt} \frac{1}{2} \sum_{k=1}^L \left ( v_{2,\alpha}^{(k)} \right )^2 \right ].
\end{equation}
According to Refs.~\onlinecite{MET_Nonham2}, if there exists a function 
$w(\rx)$ such that $\kappa(\rx) = dw(\rx)/dt$, then
\begin{equation}
\sqrt{g(\rx)} = e^{-w(\rx)},
\end{equation}
and the partition function generated by Eqs.\ (\ref{eq:eoms_an}) is
\[
\Omega(N,V,\beta)  =  \int d^{dN}a d^{dN}p d^{dNL}v_1 d^{dNL}v_2 \rho_{\rm isok},
\]
where 
\begin{equation}
\rho_{\rm isok} = e^{-\frac{\beta}{2}\sum_{k=1}^L \left ( v_{2,\alpha}^{(k)} \right )^2}  e^{-\beta U(q)} \times
\prod_{\alpha=1}^{dN}
\delta\left(v_{\alpha}^2 + {L \over L+1}\sum_{k=1}^L \left(v_{1,\alpha}^{(k)}\right)^2 - L\beta^{-1}\right),
\label{eq:isokdist}
\end{equation}
which is clearly canonical in the $dN$ coordinates $q_1,...,q_{dN}$.

Now let us consider effect of the friction and noise terms which we have so far neglected. Note that the invariant distribution generated in $v_{2,\alpha}^{(k)}$ by the Ornstein-Uhlenbeck component is Gaussian.   Since the noise process only contacts the $v_{2,\alpha}^{(k)}$ terms, it follows that the remaining part of the distribution function will be left invariant by the action of  the corresponding Fokker-Planck equation.    We may think of the stochastic differential equations as being divided into two parts:
\begin{equation} \label{SDE1}
d{\bf X} =\mathop{\vphantom{\underbrace{{\bf \Gamma} {\bf X}dt + {\bf \Sigma} d{\bf  W}}}\underbrace{{\bf f}({\bf X})dt}}_{\rm deterministic} + \underbrace{\mathop{\vphantom{{\bf f}({\bf X})dt}} {\bf \Gamma} {\bf X}dt + {\bf \Sigma} d{\bf  W}}_{\rm Ornstein-Uhlenbeck}
\end{equation}
where ${\bf f}$ corresponds to the deterministic chain system considered above, and ${\bf \Gamma}$ and ${\bf \Sigma}$ are suitable matrices which describe the Ornstein-Uhlenbeck-type (linear) stochastic differential equations in each of the $v_2$ components ($\bf W$ represents a vector of independent Wiener processes, one contacting each $v_{2,\alpha}^{(k)}$ degree of freedom).
The corresponding Fokker-Planck operator inherits an additive decomposition
\begin{equation}
\rho_t = -i{\cal L}_1^{\dagger} \rho - i{\cal L}_2^{\dagger} \rho
\end{equation}
and we have the property that, in the weak (distributional) sense
\begin{equation}
{\cal L}_1^{\dagger} \rho_{\rm isok} =0
\end{equation}
and 
\begin{equation}
{\cal L}_2^{\dagger}\left[e^{-\frac{\beta}{2}\sum_{k=1}^L \left ( v_{2,\alpha}^{(k)} \right )^2} \hat{\rho}\right] = 0,
\end{equation}
where $\hat{\rho}$ is an arbitrary, normalizable distribution in the other variables in the system (besides $v_{2,\alpha}^{(k)}$).
Clearly $({\cal L}_1^{\dagger} + {\cal L}_2^{\dagger})\rho_{\rm isok} =0$.

It therefore follows that the isokinetic density $\rho_{\rm isok}$ given above will also be preserved under the dynamics of the full system with the stochastic process incorporated.   It remains only to show that this is the unique stationary state of the SDE.

\subsection{Ergodicity}
In this subsection we sketch a proof of the ergodicity of the isokinetic model in the case of a single stochastic isokinetic thermostat on a single harmonic oscillator, which, in a certain sense, is the most difficult case (there is no internal mechanism present in the deterministic dynamics to promote mixing).
It is likely that, with more effort, the proof could be extended to
anharmonic potentials.
Theories of ergodicity for stochastic differential equations are now well developed\cite{MeynTweedie,RB06,Ha10}.   Consider a stochastic differential equation (SDE) on a smooth manifold $M$ of the form
\begin{equation}
\dd {\bf X} = {\bf a}({\bf X}) \dd t + \sum_{i=1}^m {\bf b}_i({\bf X}) \dd {\rm W}_i
\end{equation}
where $\dd {\rm W}_i$, $i=1,\ldots, m$ are the infinitesimal increments of $m$ independent Wiener processes, and ${\bf a}({\bf X})$ and ${\bf b}_i({\bf X})$, $i=1,\ldots, m$ are smooth vector fields on the tangent space $TM_{{\bf X}}$.     Note that Equation \ref{SDE1} can be put in this form by defining ${\bf a}(\bf X):={\bf f}({\bf X}) - {\bf \Gamma} {\bf X}$ and $\sum_{i=1}^m {\bf b}_i({\bf X}) \dd W_i \equiv {\bf \Gamma} \dd {\bf W}$.  As mentioned in the previous subsection, corresponding to the SDE there is a Fokker-Planck equation of the form
\begin{equation}
\rho_t = -i{\cal L}^{\dagger} \rho,
\end{equation}
where $i{\cal L}$ is the generator of the stochastic process and $-i{\cal L}^{\dagger}$ is its adjoint.  The system is ergodic, implying the
convergence of averages along almost every trajectory, provided the solution of ${\cal L}^{\dagger}\rho=0$  is unique (up to a constant scaling) and $C^{\infty}(M)$.    In the case at hand, we have a smooth stationary distribution (see the previous section) and all that remains is to check that it is unique.    We omit some details here
(see refs. \onlinecite{MSH02,LeNoTh09} for related studies) noting only that the crucial step needed to establish the regularity of the operator (and thus the uniqueness of the invariant density and hence the ergodicity of the stochastic differential equations) is to verify a certain H\"ormander condition.  The H\"ormander condition guarantees that the solutions of ${\cal L}^{\dagger} \rho=0$ are smooth (in $C^{\infty}$) and unique.

Let $\mathcal{I}({\bf b}_0:={\bf a},{\bf b}_1,\dots,{\bf b}_{m})$ denote the ideal generated by ${\bf b}_0,\ldots,{\bf b}_m$:
\[
        \mathcal{I}({\bf b}_0,{\bf b}_1,\dots,{\bf b}_{m}) = \{ {\bf b}_{k_0}, [{\bf b}_{k_0},{\bf b}_{k_1}], [{\bf b}_{k_0},[{\bf b}_{k_1},{\bf b}_{k_2}]], \dots \},
\]
where $[\cdot,\cdot]$ denotes the commutator of vector fields, $k_0$,$k_1$, $k_2$, etc.,  take values in the set $\{ 0,\dots, m\}$.

The H\"ormander condition \cite{RoWi87} to ensure a smooth invariant probability measure for this system is
\begin{equation}
        TM =\mathrm{span}\, \mathcal{I}({\bf b}_0,{\bf b}_1,\dots,{\bf b}_{m}),
\end{equation}

Intuitively, the H\"ormander condition
 implies that that at any point of our phase space, the dynamics will explore all possible directions, so noise, introduced in one component,
filters into all directions.  Note that the condition is sometimes stated   in a restricted form\cite{RB06,Ha10} which does not allow $k_0$ to have value $0$. It is more difficult in many cases to verify this restricted H\"ormander condition, the consequence of which is the smoothness of solutions to the time-dependent Fokker-Planck equation for arbitrary nonzero time $t$.   We are interested in the long-term behavior of solutions, i.e. the time-invariant solution of the stationary Fokker-Planck equation, and for this to hold, one may use the simpler H\"ormander condition given above.   Another technicality that should be addressed is the treatment of
an unbounded space;  we should consider the contractivity of the flow by constructing a Lyapunov function, as in ref. \onlinecite{MSH02}, however, with periodic boundary conditions the configuration space is bounded, and with the isokinetic constraint the velocities (momenta) and auxiliary variables $v_{1,\alpha}$ remain bounded.  All that is left to is to insure that there is no divergence in the $v_{2,\alpha}$ directions, but this is easily shown as they are controlled directly by Wiener (friction+noise) processes.


For a single harmonic oscillator with $L=1$, the isokinetic stochastic dynamics take the form
\begin{eqnarray}
\dd q & = & v \dd t \nonumber \\
\dd v & = & [-q -\lambda v] \dd t \nonumber \\
\dd v_1 & = &  [-\lambda v_1 - v_2 v_1] \dd t \nonumber \\
\dd v_2 & = & [ Q^{-1}[ v_1^2 - k_{\rm B}T] - \gamma v_2] \dd t + \sqrt{2\gamma k_BT /Q} \dd {\rm W}
\end{eqnarray}
where
\begin{equation}
\lambda = \frac{-qv - v_2 v_1^2/2}{v^2 + v_1^2/2}
\end{equation}
The dimension of the extended phase space of this model is three.
(There are 4 variables and one constraint, $v^2 + v_1^2/2 =  {\rm const}$, which defines a 3-dimensional manifold ${M}$.)  

Note that $v_1=0$ is an invariant manifold of the isokinetic equations; the solutions are confined for all time to one or the other half-space ($v_1<0$ or $v_1>0$).   Furthermore observe that the equations of motion for $q,v,v_2$ all depend on $v_1^2$ only, thus there is a symmetry between the two domains $v_1<0$ and $v_1>0$.   This symmetry also carries over to the integrand of the extended partition implying that one need only sample one or the other of the two half-spaces.

In order for the system to be ergodic, we must that verify the H\"ormander condition holds.  For our system,  ${\bf a}$ and ${\bf b}$ are the vector fields
\begin{equation}
{\bf a} = v \partial_q - (q +\lambda v)\partial_v -(\lambda v_1 + v_2 v_1)\partial_{v_1} + \left [Q^{-1}( v_1^2 - k_{\rm B}T) - \gamma v_2 \right ]\partial_{v_2},
\end{equation}
and
\begin{equation}
{\bf  b}= \sigma \partial_{v_2}
\end{equation}
where $\sigma= \sqrt{2\gamma k_{\rm B}T /Q}$.

The H\"ormander condition we require is this: at any point, the dimension of the ideal spanned by the iterated commutators of the vector fields ${\bf a}$ and ${\bf b}$,
\[
{\bf a}, {\bf b}, [{\bf a},{\bf b}], [{\bf a},[{\bf a},{\bf b}]], [{\bf b},[{\bf a},{\bf b}]],\ldots
\]
has dimension 3, that is, it spans the tangent space to the manifold $M$.   We will compute a particular selection from the commutators and show that they form a basis, specifically, we consider the vectors
\[
{\bf a}, {\bf b}, [{\bf a},{\bf b}], [{\bf a},[{\bf a},{\bf b}]],  [{\bf a}, [{\bf a},[{\bf a},{\bf b}]]].
\]
From these vectors, we will find three linearly independent ones at every point except on a one-dimensional set (the union of two lines).   We may assume $v_1\neq 0$ as mentioned above. The fact that the H\"{o}rmander condition fails on a set of dimension 1 (i.e. co-dimension 2 relative to the manifold $M$) is of no consequence as the low-dimensional set cannot restrict the volume of the region explored by stochastic paths.  (The paths easily circumnavigate this obstacle.)

Since ${\bf b}$ is a constant multiple of $[0,0,0,1]^T$, then it is clear that ${\bf a}$ and ${\bf b}$ are linearly independent as long as one of the first three components of ${\bf a}$ is nonzero.   Having all these zero implies $q=0, v=0$, in which case $v_1$ is fixed by the isokinetic constraint, which defines two lines of degenerate points  given by $q=0, v=0$, $v_1 = \pm \sqrt{3k_{\rm B}T}$, with $v_2$ arbitrary; denote the union of these two lines by $M_0$.  

Now the commutator ${\bf a}$ and ${\bf b}$ is given by
\begin{eqnarray}
\left [ {\bf a},{\bf b} \right ] & =
 ( \sigma v \frac{\partial \lambda}{\partial v_2} )\partial_{v}
 +\sigma v_1 \left(1 + \frac{\partial \lambda}{\partial v_2}\right) \partial_{v_1}
+ \sigma  \gamma \partial_{v_2} \nonumber \\
&  = \sigma  v \frac{\partial \lambda}{\partial v_2} \partial_{v}
+ v_1 \left(1 + \frac{\partial \lambda}{\partial v_2}\right) \partial_{v_1}+ \sigma  \gamma \partial_{v_2}
\end{eqnarray}

Since ${\bf b}=\sigma {\bf e}_4$, we only need to show that, except possibly on $M_0$, the vectors defined by the first three components of ${\bf a}$ and $[{\bf a},{\bf b}]$ are linearly independent.  These are
\begin{equation}
{\bf u}_1 =  v{\partial_q} -(q+\lambda v)\partial_{v} -(\lambda v_1 +v_1 v_2)\partial_{v_1},
\end{equation}
and
\begin{equation}
{\bf u}_2 = \sigma  v \frac{\partial \lambda}{\partial v_2} \partial_{v}
+ v_1 \left(1 + \frac{\partial \lambda}{\partial v_2}\right) \partial_{v_1}.
\end{equation}
After expanding the derivatives of $\lambda$ we have
\begin{equation}
{\bf u}_1 = v \partial_{q}  + v_1^2 \left ( \frac{-q + v v_2 }{D}\right )\partial_{v} -2 v_1 v \left ( \frac{-q + v v_2}{D} \right ) \partial_{v_1},
\end{equation}
and
\begin{equation}
{\bf u}_2 = -\sigma (v_1^2v/D) \partial_{v}+ (2\sigma v_1 v^2/D) \partial_{v_1},
\end{equation}
where $D=2 v^2 + v_1^2$.

If $q\neq 0$ and $v\neq 0$, then clearly ${\bf u}_1$ and ${\bf u}_2$ are linearly independent. However, if $v=0$ then ${\bf u}_2=0$.

For this reason we must compute an additional commutator $[{\bf a},[{\bf a},{\bf b}]]$.   Projecting to the first three components (since we already have that ${\bf e}_4$ is in our subspace) results in
\begin{equation}
{\bf u}_3 =\sigma \left (\frac{v_1^2 v}{D} \right) \partial_q -\sigma\left ( \frac{2 \gamma v^3 +\gamma v_1^2 v - v_1^2 q}{D^2} \right ) \partial_{v}
+ 2 \sigma v_1 v \left ( \frac{2 \gamma v^3 + \gamma v_1^2 v-v_1^2 q}{D^2}\right ) \partial_{v_1}.
\end{equation}
Again, we substitute $v=0$ and find that
\begin{equation}
\left . {\bf u}_3\right |_{v=0} = (\sigma v_1^2 q/D^2)\partial_{v} .
\end{equation}
Unfortunately, this is parallel to ${\bf u}_1$ (for $v=0$), and hence another commutator is required.  Computing $[{\bf a}, [{\bf a},[{\bf a},{\bf b}]]]$ and projecting to the first three components yields
\begin{eqnarray}
{\bf u}_4 & = & \sigma \left (\frac{2\gamma v^3-4v^3 v_2+4v^2 q+v_1^2 v v_2+\gamma v_1^2 v-2 v_1^2 q}{D^2} \right) \partial_q
\nonumber \\
\nonumber \\
& - & \sigma v_1^2 \left ( \frac{\eta}{D^3} \right ) \partial_{v}
+ 2 \sigma v_1 v \left ( \frac{\eta}{D^3}\right ) \partial_{v_1}.
\end{eqnarray}
where
\begin{eqnarray}
\eta  & = & 4 v^5 \gamma^2-8 v^5 Q^{-1} v_1^2-4 v_1^2 v^3+4 v^3 \gamma^2 v_1^2-4 v^3 v_1^4 Q^{-1}\nonumber \\
& - & 2\gamma v_1^2 v^2 q+4 v^2 q v_2 v_1^2-4 v_1^2 q^2 v-2 v_1^4 v+v \gamma^2 v_1^4-q \gamma v_1^4+v_1^4 v_2 q.
\end{eqnarray}

Now along $v=0$ we find that ${\bf u}_4$ simplifies to
\begin{eqnarray}
\left . {\bf u}_4\right |_{v=0} & = &  -2 \sigma q \partial_{q} -\sigma \left ( \frac{ -q \gamma v_1^4+v_1^4 v_2 q}{v_1^4}\right )\partial_{v} 
\nonumber \\
\nonumber \\
& = & -2 \sigma q \partial_{q} -\sigma q (-\gamma + v_2 )\partial_{v}
\end{eqnarray}
Finally we see that this and ${\bf u}_1$ form a linearly independent set if $q\neq 0$. (Of course if $v=q=0$ then we are on $M_0$.)

Thus we see that, off of the low-dimensional set $M_0$, the H\"ormander condition is verified for the stochastic isokinetic method applied to the harmonic oscillator.   We therefore conclude that the invariant measure of the stochastic isokinetic system (in the case of the harmonic oscillator) is unique, and thus that the process is ergodic.

\section{Multiple time step integration algorithm}
\label{sec:integrator}

In this section, 
we derive a multiple time-step (MTS) integrator for the equations of motion (\ref{eq:isokink}),  based on the reference system propagator algorithm (RESPA)
introduced by Tuckerman {\it et al.}~\cite{Tuckerman92}.  In this derivation, in order to simplify the notation, we will drop the
$\alpha$ index, which labels the degrees of freedom in the system.  However, it must be kept in mind that the
integrator obtained should be applied to {\it each} degree of freedom in the system.

The derivation begins
with the Liouville operator for the equations of motion given by
\begin{equation}
iL = iL_q + iL_v + iL_N + iL_{\rm OU}
\end{equation}
where
\begin{eqnarray}
iL_q & = & v{\partial \over \partial q} \nonumber  \\
\nonumber \\
iL_v & = & \left({F \over m}-\lambda_F v\right){\partial \over \partial v} - 
\lambda_F \sum_{k=1}^L v_1^{(k)}{\partial \over \partial v_1^{(k)}}
\nonumber \\
\nonumber \\
iL_N & = & -\lambda_N v {\partial \over \partial v} - \lambda_N 
\sum_{k=1}^L v_1^{(k)} {\partial \over \partial v_1^{(k)}}
-\sum_{k=1}^L v_2^{(k)} v_1^{(k)} {\partial \over \partial v_1^{(k)}} 
+ \sum_{k=1}^L {G(v_1^{(k)})\over Q_2}{\partial \over \partial v_2^{(k)}}
\label{eq:lville}
\end{eqnarray}
and $iL_{\rm OU}$ corresponds to the Ornstein-Uhlenbeck-type stochastic process
applied to $v_2^{(k)}$.  A numerical integrator is derived via an MTS
factorization of the classical propagator $\exp(iL\Delta t)$
based on the Trotter theorem, where $\Delta t$ is a time
step appropriate for the slowest motion.  
In the derivation to 
follow, we will build on a basic factorization scheme for Langevin dynamics recently
studied in depth by Matthews and Leimkuhler~\cite{LeMa2013}, in particular separating and treating exactly the Ornstein-Uhlenbeck term.  
The multiplier in Eq.\ (\ref{eq:lambdak}) contains two contributions, which we
express as $\lambda = \lambda_F + \lambda_N$, referring to the 
contributions from the force $F$ and the Nos\'e-like coupling to the
extended phase-space variables $v_1^{(k)}$.  For standard
Langevin dynamics, obtained from Eq.\ (\ref{eq:lville}) by setting
$iL_N=0$ and $\lambda_F = \lambda_N=0$ so that
$iL_v = (F/m)(\partial/\partial v)$, the factorization takes the form
\begin{equation}
e^{iL\Delta t} = e^{iL_v\Delta t/2} e^{iL_q \Delta t/2} e^{iL_{\rm OU}\Delta t} e^{iL_q \Delta t/2} e^{iL_v\Delta t/2} 
\label{eq:ML}
\end{equation}
Extending this to the stochastic isokinetic Liouville operator
of Eq.\ (\ref{eq:lville}), the corresponding single time-step
integrator would take the form
\begin{equation}
e^{iL\Delta t} = e^{iL_N\Delta t/2}e^{iL_v\Delta t/2} e^{iL_q \Delta t/2} e^{iL_{\rm OU}\Delta t} e^{iL_q \Delta t/2} e^{iL_v\Delta t/2} 
e^{iL_N\Delta t/2}
\label{eq:ML_SIN}
\end{equation}
Eq.\ (\ref{eq:ML_SIN}) could be employed as a starting point for the derivation of a 
robust numerical scheme by applying it together with the integrators of 
Refs.~\onlinecite{TUCK_ISOK1,Minary_PRL}.

In order to derive an MTS
algorithm, suppose the force contains
a fast and a slow component:  $F = F_f + F_s$.  
With this division, the Liouville operator can be expressed as
\begin{equation}
iL = iL_q + iL_v^{(f)} + iL_v^{(s)} + iL_N + iL_{\rm OU}
\end{equation}
where the contributions $iL_v^{(f)}$ and $iL_v^{(s)}$ are determined by
the fast and slow force components and their contributions
to the Lagrange multiplier:
\begin{eqnarray}
iL_v^{(f)} & = & \left({F_f \over m}-\lambda_F^{(f)} v\right){\partial \over \partial v} - 
\lambda_F^{(f)} \sum_{k=1}^L v_1^{(k)}{\partial \over \partial v_1^{(k)}}
\nonumber \\
\nonumber \\
iL_v^{(s)} & = & \left({F_s \over m}-\lambda_F^{(s)} v\right){\partial \over \partial v} - 
\lambda_F^{(s)} \sum_{k=1}^L v_1^{(k)}{\partial \over \partial v_1^{(k)}}
\label{eq:lville_vfs}
\end{eqnarray}
The corresponding Lagrange multiplier contributions are
\begin{eqnarray}
\lambda_F^{(f)} & = & {v F_f \over \Lambda}
\nonumber \\
\nonumber \\
\lambda_F^{(s)} & = & {v F_s \over \Lambda}
\nonumber \\
\nonumber \\
\lambda_N & = & -{(L/L+1) \sum_{j=1}^L Q_1\left(v_1^{(j)}\right)^2 v_2^{(j)} \over \Lambda}
\end{eqnarray}
and $\Lambda = mv^2 + (L/L+1)\sum_k Q_{1}\left(v_1^{(k)}\right)^2 = L\beta^{-1}$.

RESPA integrators for extended-system thermostatted MD
equations of motion are of two types described in 
Ref.~\onlinecite{Martyna96} depending on the placement of the evolution
step of the extended phase-space variables in the algorithm.
When these steps are at the beginning and end of the overall
integration step, the approach is called an extended-system
``outer'' RESPA or XO-RESPA scheme.  When these steps are
carried out with the evolution of the fastest motion,
the scheme is called an extended-system inner RESPA
or XI-RESPA scheme.  These two schemes involve different
factorizations of the classical propagator.  Let
$\dt$ and $\dti$ be the time steps associated with the
slow and fast force components, respectively.  Then,
the proposed XO-RESPA factorization can be written as
\begin{eqnarray}
& & \exp(iL\dt) = \exp\left(iL_N {\dt \over 2}\right)
\nonumber \\
\nonumber \\
& & \;\;\;\;\;\;\;\;\;\;\times 
\exp\left\{{\dti \over 2}\left[\left({F_f + nF_s \over m}
- \left(\lambda_F^{(f)} + n\lambda_F^{(s)}\right)\right){\partial \over
\partial v} - \left(\lambda_F^{(f)} + n\lambda_F^{(s)}\right)
\sum_{k=1}^L v_1^{(k)}{\partial \over \partial v_1^{(k)}}\right]\right\}
\nonumber \\
\nonumber \\
& & \;\;\;\;\;\;\;\;\;\;\times \exp\left({\dti \over 2}v{\partial \over \partial q}\right)
\exp\left(iL_{{\rm OU}}\dti\right)
\exp\left({\dti \over 2}v{\partial \over \partial q}\right)
\nonumber \\
\nonumber \\
& & \;\;\;\;\;\;\;\;\;\;\times 
\exp\left\{{\dti \over 2}\left[\left({F_f \over m}
- \lambda_F^{(f)}\right){\partial \over\partial v} 
- \lambda_F^{(f)}
\sum_{k=1}^L v_1^{(k)}{\partial \over \partial v_1^{(k)}}\right]\right\}
\nonumber \\
\nonumber \\
& & \;\;\;\;\;\;\;\;\;\;\times \left[
\exp\left\{{\dti \over 2}\left[\left({F_f \over m}
- \lambda_F^{(f)}\right){\partial \over\partial v} 
- \lambda_F^{(f)}
\sum_{k=1}^L v_1^{(k)}{\partial \over \partial v_1^{(k)}}\right]\right\}
\right.
\nonumber \\
\nonumber \\
& & \;\;\;\;\;\;\;\;\;\;\times \exp\left({\dti \over 2}v{\partial \over \partial q}\right)
\exp\left(iL_{{\rm OU}}\dti\right)
\exp\left({\dti \over 2}v{\partial \over \partial q}\right)
\nonumber \\
\nonumber \\
& & \left.\;\;\;\;\;\;\;\;\;\;\times
\exp\left\{{\dti \over 2}\left[\left({F_f \over m}
- \lambda_F^{(f)}\right){\partial \over\partial v} 
- \lambda_F^{(f)}
\sum_{k=1}^L v_1^{(k)}{\partial \over \partial v_1^{(k)}}\right]\right\}\right]^{n-2}
\nonumber \\
\nonumber \\
& & \;\;\;\;\;\;\;\;\;\;\times
\exp\left\{{\dti \over 2}\left[\left({F_f \over m}
- \lambda_F^{(f)}\right){\partial \over\partial v} 
- \lambda_F^{(f)}
\sum_{k=1}^L v_1^{(k)}{\partial \over \partial v_1^{(k)}}\right]\right\}
\nonumber \\
\nonumber \\
& & \;\;\;\;\;\;\;\;\;\;\times 
\exp\left({\dti \over 2}v{\partial \over \partial q}\right)
\exp\left(iL_{{\rm OU}}\dti\right)
\exp\left({\dti \over 2}v{\partial \over \partial q}\right)
\nonumber \\
\nonumber \\
& & \;\;\;\;\;\;\;\;\;\;\times
\exp\left\{{\dti \over 2}\left[\left({F_f + nF_s \over m}
- \left(\lambda_F^{(f)} + n\lambda_F^{(s)}\right)\right){\partial \over
\partial v} - \left(\lambda_F^{(f)} + n\lambda_F^{(s)}\right)
\sum_{k=1}^L v_1^{(k)}{\partial \over \partial v_1^{(k)}}\right]\right\}
\nonumber \\
\nonumber \\
& & \;\;\;\;\;\;\;\;\;\;\times 
\exp\left(iL_N {\dt \over 2}\right)
\label{eq:xo-respa}
\end{eqnarray}
The form of this factorization is largely formal in its
construction.  In practice, there is {\it no need} to split the evolution up into
first, $n-2$ intermediate, and last steps.  Rather, in the first and
last iterations of the RESPA loop, the force is taken to be
$F_f + nF_s$ while for the remaining $n-2$ iterations, it is simply $F_f$.
Also, within this scheme, evolution produced by the stochastic force term, which only affects
$v_2^{(k)}$ can be derived using the It$\bar{\rm o}$ calculus and
is given by
\begin{equation}
e^{iL_{\rm OU} t}v_2^{(k)} = v_2^{(k)}(0) e^{-\gamma t} + \sigma R(t) \sqrt{{1 - e^{-2\gamma t} \over 2\gamma}}
\label{eq:W_op}
\end{equation}
where $R$ is the random force at time $t$.  

Note that in XO-RESPA, the purely extended system part involving
$iL_N$ is evaluated once at the beginning and once at the 
end of the step using a time step $\dt$.  For XI-RESPA, this
operator is evaluated every small time step using the factorization:
\begin{eqnarray}
& & \exp(iL\dt) = \exp\left(iL_N {\dti \over 2}\right)
\nonumber \\
\nonumber \\
& & \;\;\;\;\;\;\;\;\;\;\times
\exp\left\{{\dti \over 2}\left[\left({F_f + nF_s \over m}
- \left(\lambda_F^{(f)} + n\lambda_F^{(s)}\right)\right){\partial \over
\partial v} - \left(\lambda_F^{(f)} + n\lambda_F^{(s)}\right)
\sum_{k=1}^L v_1^{(k)}{\partial \over \partial v_1^{(k)}}\right]\right\}
\nonumber \\
\nonumber \\
& & \;\;\;\;\;\;\;\;\;\;\times \exp\left({\dti \over 2}v{\partial \over \partial q}\right)
\exp\left(iL_{{\rm W}}\dti\right)
\exp\left({\dti \over 2}v{\partial \over \partial q}\right)
\nonumber \\
\nonumber \\
& & \;\;\;\;\;\;\;\;\;\;\times 
\exp\left\{{\dti \over 2}\left[\left({F_f \over m}
- \lambda_F^{(f)}\right){\partial \over\partial v} 
- \lambda_F^{(f)}
\sum_{k=1}^L v_1^{(k)}{\partial \over \partial v_1^{(k)}}\right]\right\}
\nonumber \\
\nonumber \\
& & \;\;\;\;\;\;\;\;\;\;\times \left[\exp\left(iL_N {\dti \over 2}\right)
\exp\left\{{\dti \over 2}\left[\left({F_f \over m}
- \lambda_F^{(f)}\right){\partial \over\partial v} 
- \lambda_F^{(f)}
\sum_{k=1}^L v_1^{(k)}{\partial \over \partial v_1^{(k)}}\right]\right\}
\right.
\nonumber \\
\nonumber \\
& & \;\;\;\;\;\;\;\;\;\;\times \exp\left({\dti \over 2}v{\partial \over \partial q}\right)
\exp\left(iL_{{\rm W}}\dti\right)
\exp\left({\dti \over 2}v{\partial \over \partial q}\right)
\nonumber \\
\nonumber \\
& & \left.\;\;\;\;\;\;\;\;\;\;\times
\exp\left\{{\dti \over 2}\left[\left({F_f \over m}
- \lambda_F^{(f)}\right){\partial \over\partial v} 
- \lambda_F^{(f)}
\sum_{k=1}^L v_1^{(k)}{\partial \over \partial v_1^{(k)}}\right]\right\}
\exp\left(iL_N {\dti \over 2}\right)\right]^{n-2}
\nonumber \\
\nonumber \\
& & \;\;\;\;\;\;\;\;\;\;\times
\exp\left\{{\dti \over 2}\left[\left({F_f \over m}
- \lambda_F^{(f)}\right){\partial \over\partial v} 
- \lambda_F^{(f)}
\sum_{k=1}^L v_1^{(k)}{\partial \over \partial v_1^{(k)}}\right]\right\}
\nonumber \\
\nonumber \\
& & \;\;\;\;\;\;\;\;\;\;\times 
\exp\left({\dti \over 2}v{\partial \over \partial q}\right)
\exp\left(iL_{{\rm W}}\dti\right)
\exp\left({\dti \over 2}v{\partial \over \partial q}\right)
\nonumber \\
\nonumber \\
& & \;\;\;\;\;\;\;\;\;\;\times
\exp\left\{{\dti \over 2}\left[\left({F_f + nF_s \over m}
- \left(\lambda_F^{(f)} + n\lambda_F^{(s)}\right)\right){\partial \over
\partial v} - \left(\lambda_F^{(f)} + n\lambda_F^{(s)}\right)
\sum_{k=1}^L v_1^{(k)}{\partial \over \partial v_1^{(k)}}\right]\right\}
\nonumber \\
\nonumber \\
& & \;\;\;\;\;\;\;\;\;\;\times
\exp\left(iL_N {\dti \over 2}\right)
\label{eq:xi-respa}
\end{eqnarray}
The action of $\exp(iL_{\rm OU}t))$ on the $v_2^{(k)}$ is shown in 
Eq.\ (\ref{eq:W_op}), and the action of the operator
$\exp(tv\partial/\partial q)$ on $q$ is a simple
translation $q \rightarrow q + vt$.  The action of the
reamining operators is discussed below.

\subsection{Solution for $\exp(iL_vt)$}

The force appearing in the propagator $\exp(iL_v t)$ is either $F_f$
or $F_f + nF_s$ depending on the step of the RESPA loop.
As both $F_f$ and $F_s$ are both independent of $v$,
we can simply solve the problem for a general $F$, which
is either $F_f$ or $F_f + nF_s$, depending on the operator
that is being applied.  For any general $F$, action
of the operator $\exp(iL_v t)$ is equivalent to the solution
of the differential
equations 
\begin{eqnarray}
\dot{v} & = & {F \over m} - {v^2 F \over \Lambda} \nonumber \\
\nonumber \\
\dot{v}_{1}^{(k)} & = & -{vF \over \Lambda}v_1^{(k)}
\end{eqnarray}
where $\Lambda = L\beta^{-1}$.  In these equations, $F$ is treated as a constant, 
and the equations must be solved for an arbitrary initial
condition $v(0)$, $v_1^{(k)}(0)$.  These equations are nonlinear, however,
an analytical solution is actually available for them.
Following the procedure of Ref.~\onlinecite{TUCK_ISOK1}, we write the differential equations in 
the form
\begin{eqnarray}
\dot{v} & = & {F \over m} - \dot{h}v \nonumber \\
\nonumber \\
\dot{v}_{1}^{(k)} & = & -\dot{h}v_1^{(k)}
\end{eqnarray}
where $\dot{h} = v(t) F/\Lambda$.  We then assume a solution of the form
\begin{eqnarray}
v(t) & = & {v(0) + (F/m)s(t) \over \dot{s}(t)}
\nonumber \\
\nonumber \\
v_1^{(k)}(t) & = & {v_1^{(k)}(0) \over \dot{s}(t)}
\label{eq:ansatz}
\end{eqnarray}
where $s(t)$ is a function to be determined.
Differentiating the ansatz for $v(t)$ with respect to time, we obtain
\begin{equation}
\dot{v} = {F \over m} - {\ddot{s} \over \dot{s}}v(t)
\end{equation}
so that $\ddot{s}/\dot{s} = \dot{h}$.  Given this relation, we see immediately that
\begin{eqnarray}
\dot{s}(t) & = & \exp\left[\int_0^t \dot{h}(t')dt'\right] = e^{h(t)}e^{-h(0)}
\nonumber \\
\nonumber \\
s(t) & = & \int_0^t dt' e^{h(t')}
\end{eqnarray}
so that $s(0) = 0$, and $\dot{s}(0) = 1$.  The equation that $s(t)$ must satisfy is
\begin{equation}
{\ddot{s} \over \dot{s}} = \dot{h} = {F \over \Lambda}\left[{v(0) + (F/m)s(t) \over \dot{s}}\right]
\end{equation}
or
\begin{equation}
\ddot{s} - {F^2 \over m\Lambda}s - {F \over \Lambda}v(0) = 0
\end{equation}
This equation can be easily
solved using Laplace transforms to yield
\begin{equation}
s(t) = {1 \over \sqrt{b}}\sinh\left(\sqrt{b}t\right) + 
{a \over b}\left(\cosh\left(\sqrt{b}t\right) - 1\right)
\end{equation}
where $a = Fv(0)/\Lambda$ and $b = F^2/m\Lambda$.
With this and the corresponding expression for $\dot{s}$
\begin{equation}
\dot{s}(t) = \sinh\left(\sqrt{b} t\right) + {a \over \sqrt{b}}\cosh\left(\sqrt{b}t\right)
\end{equation}
Thus, the application of $\exp(iL_v\dti/2)$ yields the following evolution step:
\begin{eqnarray}
v\left({\dti \over 2}\right) & = & {v(0) + (F/m)s(\dti/2) \over \dot{s}(\dti/2)}
\nonumber \\
v_1^{(k)}\left({\dti \over 2}\right) & = & {v_1^{(k)}(0) \over \dot{s}(\dti/2)}
\label{eq:soln_v}
\end{eqnarray}
Note that if $a$ and $b$ are close to zero, the functions $s(t)$ and $\dot{s}(t)$ 
become 0/0 and should be evaluated
as a power series in $t$.  In practice, we have found that a fourth-order power
series is sufficient when $\dti\sqrt{b}/2 < 10^{-5}$.

\subsection{Solution for $\exp(iL_N t)$}

The operator $\exp(iL_N t)$ combines a Nos\'e-Hoover type evolution
with a part of the isokinetic constraint.  Thus, 
in order to solve this part of the problem, we
first introduce a Suzuki-Yoshida 
factorization~\cite{Suzuki_85,Yoshida_90,Suzuki_91,Suzuki_92} and write
\begin{equation}
e^{iL_N \tau/2} = \prod_{j=1}^{n_{sy}}\prod_{i=1}^{n_{\rm res}} e^{iL_N w_j\tau/2n_{\rm res}}
\end{equation}
where $\tau = \dt$ or $\tau=\dti$, depending on the choice of
XO-RESPA or XI-RESPA.  Here,
$w_j$ are the Suzuki-Yoshida weights, and $n_{\rm res}$ is associated
with a second RESPA decomposition for this operator only
and should not be confused with the RESPA decomposition being applied
to the fast and slow forces.  If, for example, $n_{\rm sy} = 3$, then
the weights are $w_1 = w_3 = 1/(2-2^{1/3})$, and $w_2 = 1-w_1-w_3$.
Before we can proceed, we must further
subdivide the operator $\exp(iL_Nw_i\tau/2n_{\rm res})$,
as we cannot solve its full action
explicitly.  Thus, we write
\begin{equation}
iL_N = iL_{N,1} + iL_{N,2}
\end{equation}
where 
\begin{eqnarray}
iL_{N,1} & = & -\lambda_N v{\partial \over \partial v} - \lambda_N 
\sum_{k=1}^L v_1^{(k)} {\partial \over \partial v_1^{(k)}}
- \sum_{k=1}^L v_2^{(k)} v_1^{(k)} {\partial \over \partial v_1^{(k)}}
\nonumber \\
iL_{N,2} & = & \sum_{k=1}^L {G(v_1^{(k)}) \over Q_2}{\partial \over \partial v_2^{(k)}}
\end{eqnarray}
and we write
\begin{equation}
e^{iL_N w_j\tau/2n_{\rm res}} = e^{iL_{N,2} w_j\tau/4n_{\rm res}} 
e^{iL_{N,1}w_j\tau/2n_{\rm res}}
e^{iL_{N,2} w_j\tau/4n_{\rm res}} 
\end{equation}
The operator $\exp(iL_{N,2}w_j\tau/4n_{\rm res})$ is just a simple translation operator.

The general evolution produced by the operator $\exp(iL_{N,1} \tau/2n_{\rm res})$ 
is equivalent to the solution of the 
differential equations
\begin{eqnarray}
\dot{v} & = & {L \over (L+1)\Lambda}\left(\sum_{j=1}^L v_2^{(j)} \left(v_1^{(j)}\right)^2\right) v
\nonumber \\
\nonumber \\
\dot{v}_{1}^{(k)} & = & {L \over (L+1)\Lambda} \left(\sum_{j=1}^L
v_2^{(j)}\left(v_1^{(j)}\right)^2\right) v_1^{(k)} - v_2^{(k)} v_1^{(k)}
\end{eqnarray}
which are solved holding the $v_2^{(k)}$ fixed,
and evaluation of the solution at $t = \tau/2n_{\rm res}$.
The equations must be solved for an arbitrary initial condition
$v(0)$ and $v_1^{(k)}(0)$.
Once again, although the equations are nonlinear, 
an analytical solution can be obtained by solving the equation for $v_1^{(k)}$ by direct separation,
and then substituting the result into the equation for $v$.  This procedure yields
\begin{eqnarray}
v(t) & = &  v(0) H(t)
\nonumber \\
v_1^{(k)}(t) & = & v_1^{(k)}(0)H(t)e^{-v_2^{(k)} t}
\end{eqnarray}
where 
\begin{equation}
H(t) = \sqrt{{\Lambda \over mv^2(0) + {L \over L+1} \sum_{j=1}^LQ_1v_1^{(j)}(0)^2 e^{-2v_2^{(j)} t}}}
\end{equation}
Setting $t = \tau/2n_{\rm res}$, these become
\begin{eqnarray}
v\left({\tau \over 2n_{\rm res}}\right) & = &  v(0) H(\tau/2n_{\rm res})
\nonumber \\
\nonumber \\
v_1^{(k)}\left({\tau \over 2n_{\rm res}}\right) & = & v_1^{(k)}(0)H(\tau/2n_{\rm res})e^{-v_2^{(k)} \tau/2n_{\rm res}}
\nonumber \\
\nonumber \\
H\left({\tau \over 2n_{\rm res}}\right) & = & \sqrt{{\Lambda \over mv^2(0) + {L \over L+1} \sum_{j=1}^LQ_1
\left(v_1^{(j)}\right)^2(0) e^{-2v_2^{(j)} \tau/2n_{\rm res}}}}
\end{eqnarray}
With the action of each operator specified, the entire MTS integrator is now explicit.

The stochastic-isokinetic Nos\'e-Hoover RESPA-based MTS integrator derived in this section is
termed SIN(R) for short.  
Compared to the scheme of Ref.~\onlinecite{Minary_PRL}, the SIN(R) method is, despite 
appearances, considerably simpler to implement.

\section{Numerical examples}
\label{sec:applications}

In this section we provide numerical results which further inform the theoretical discussions of the previous sections.    
In the previous section, we described 
a numerical scheme, referred to as SIN(R), which is applicable to the general $N$-degree of freedom system and arbitrary $L$.

\subsection{Weakly perturbed harmonic oscillator}

\begin{figure}[htpb]
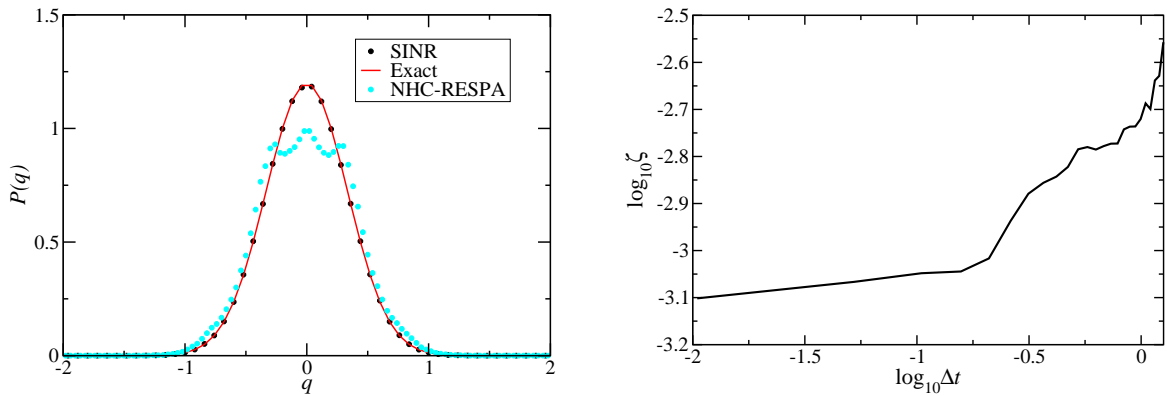

\vspace{0.25in}
\includegraphics[scale=0.3]{SINR_qdist_qHO.eps} \hspace{0.25in}
\includegraphics[scale=0.29]{log_err.eps}
\caption{(Left):  Probability distribution $P(q)$ for the
quartic potential in Eq.\ (\ref{eq:quart_pot}).  
Exact distribution (solid), SINR (black dots), 
Nos\'e-Hoover chain RESPA (cyan dots). (Right):  Logarithm of 
the L$^1$ error vs. the logarithm of the large time step
keeping the small time step fixed.}
\label{fig:quart_pot}
\end{figure}

\begin{figure}[htpb]
\includegraphics[scale=0.3, angle = -90]{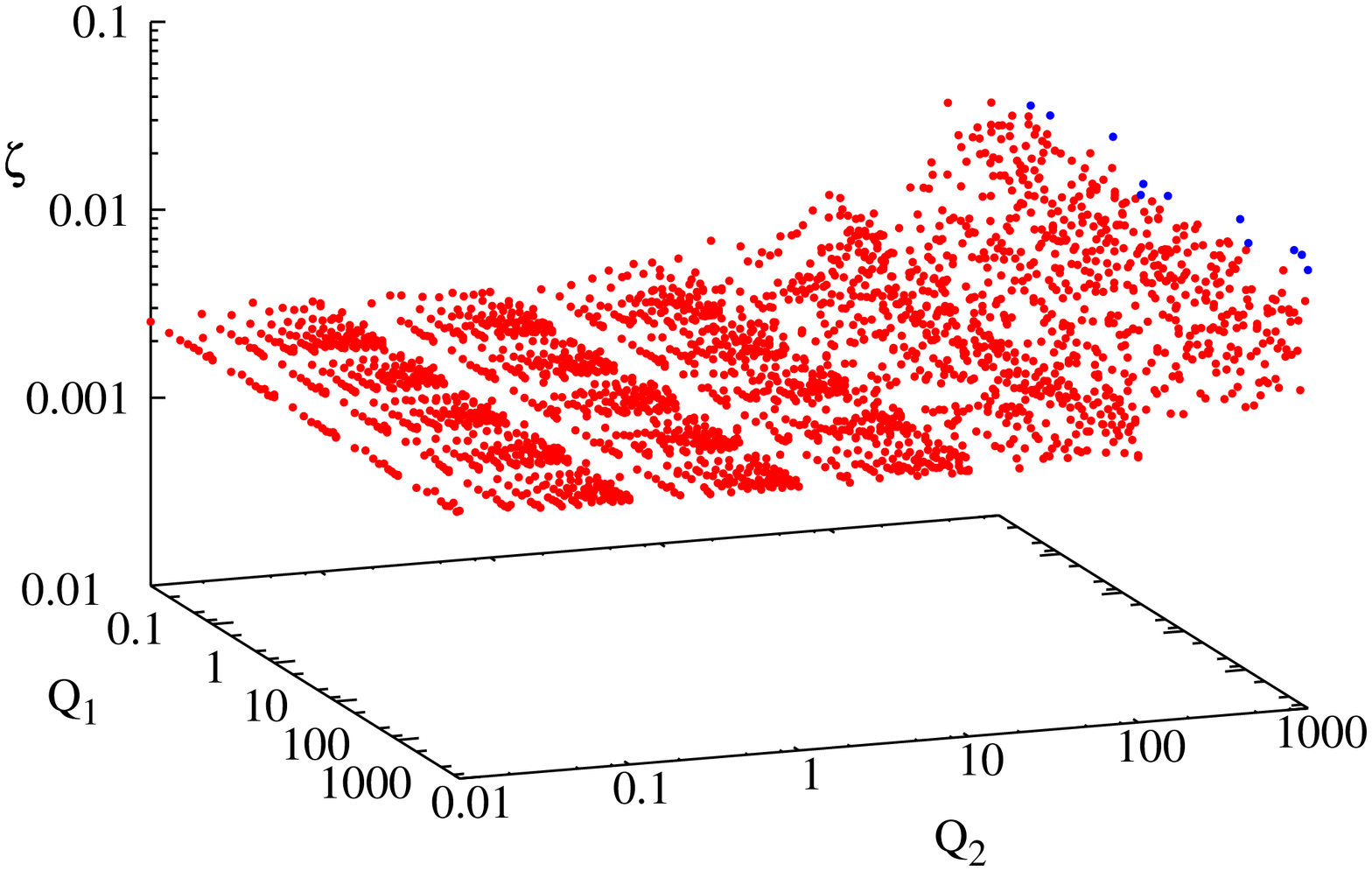}
\includegraphics[scale=0.3, angle = -90]{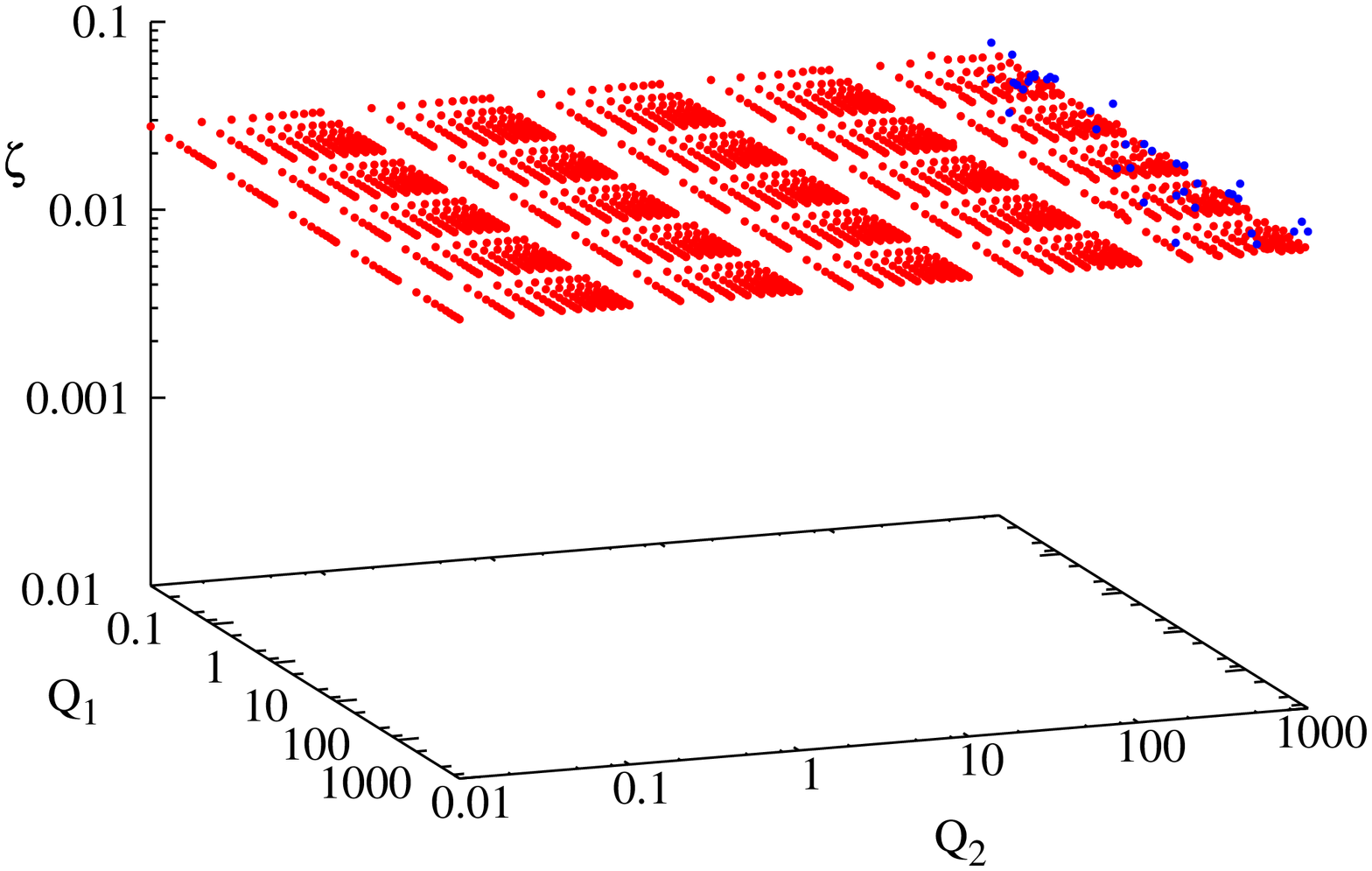} \\
\includegraphics[scale=0.3, angle = -90]{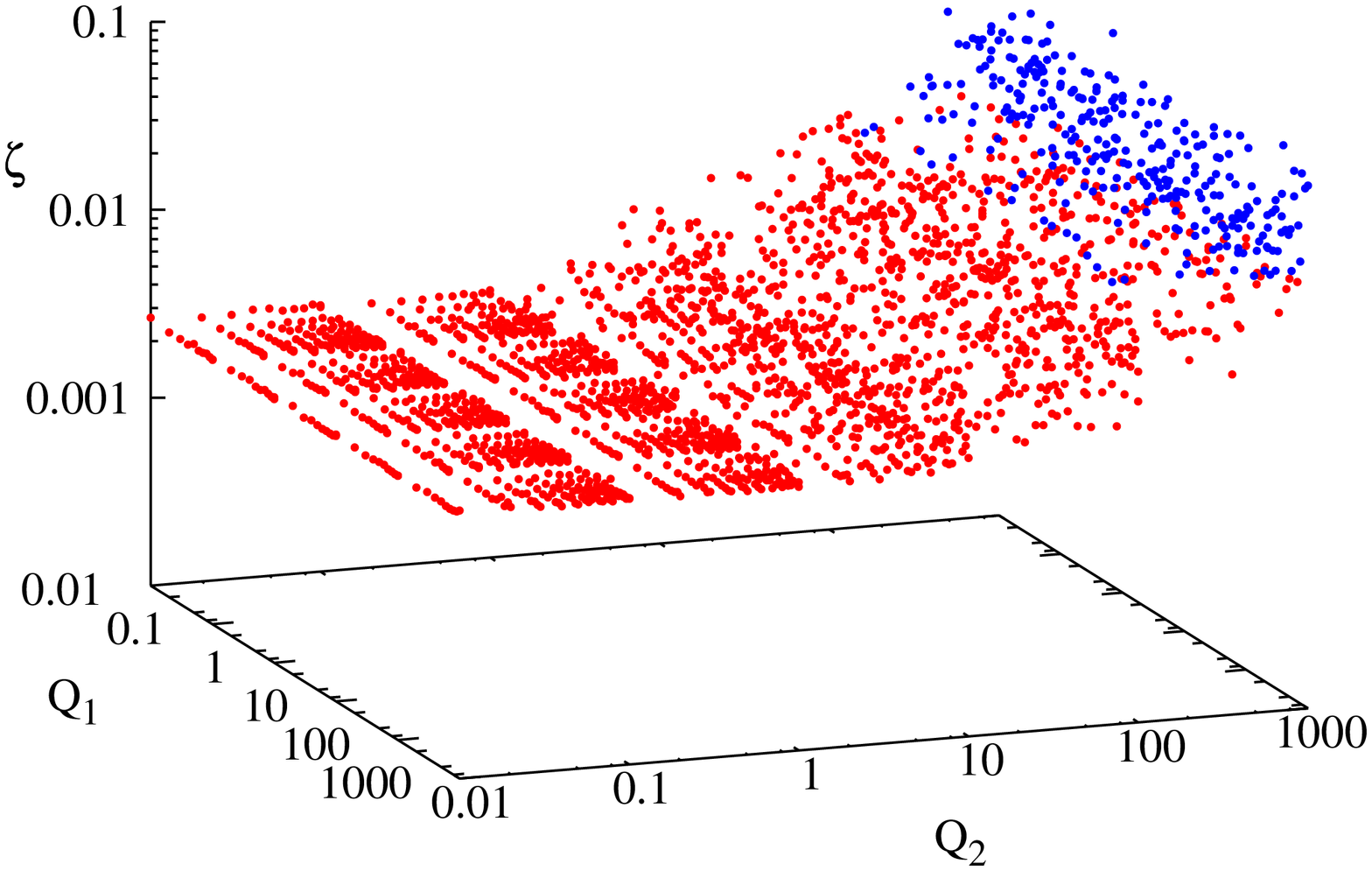} 
\includegraphics[scale=0.3, angle = -90]{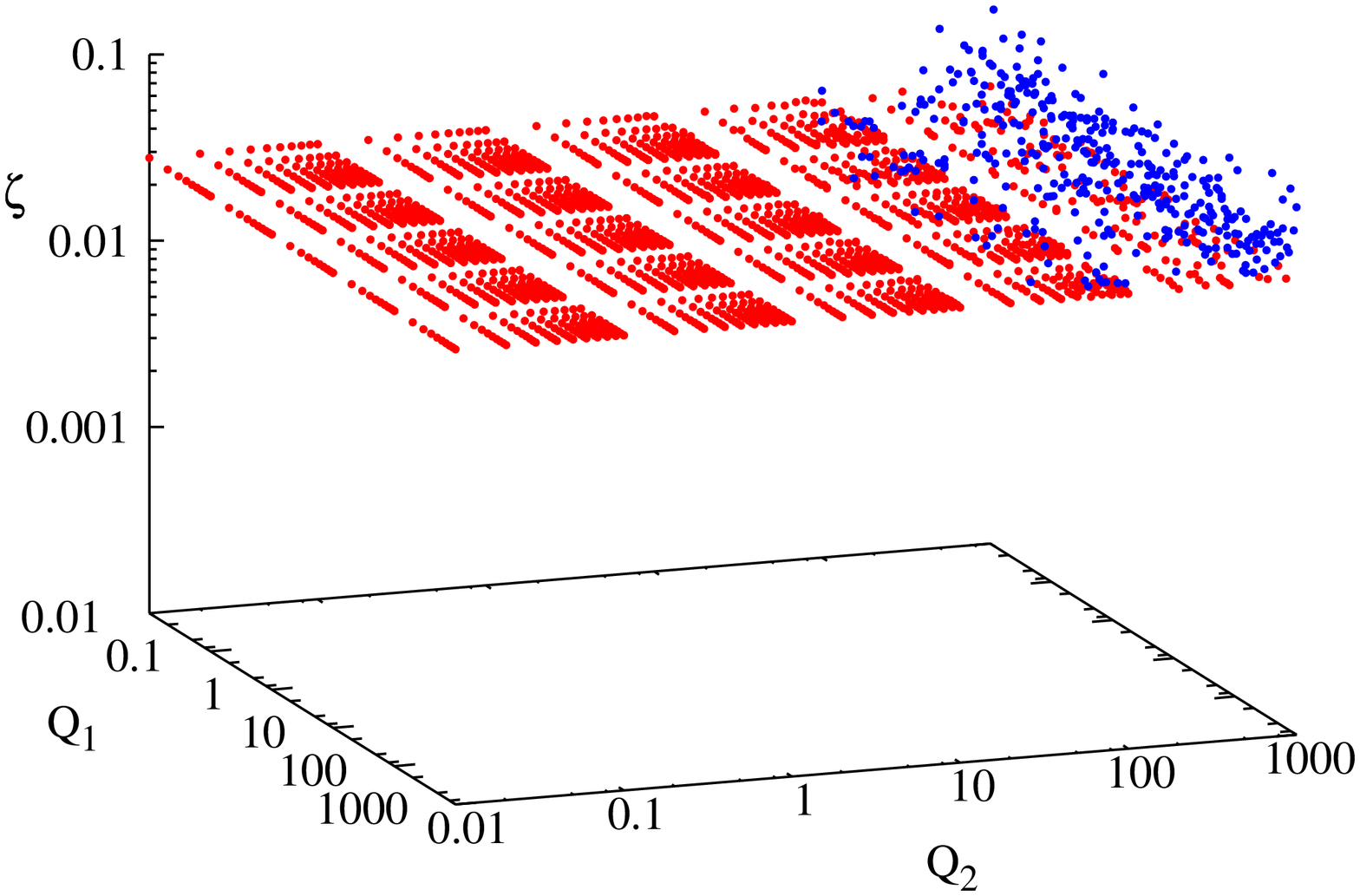} \\
\includegraphics[scale=0.3, angle = -90]{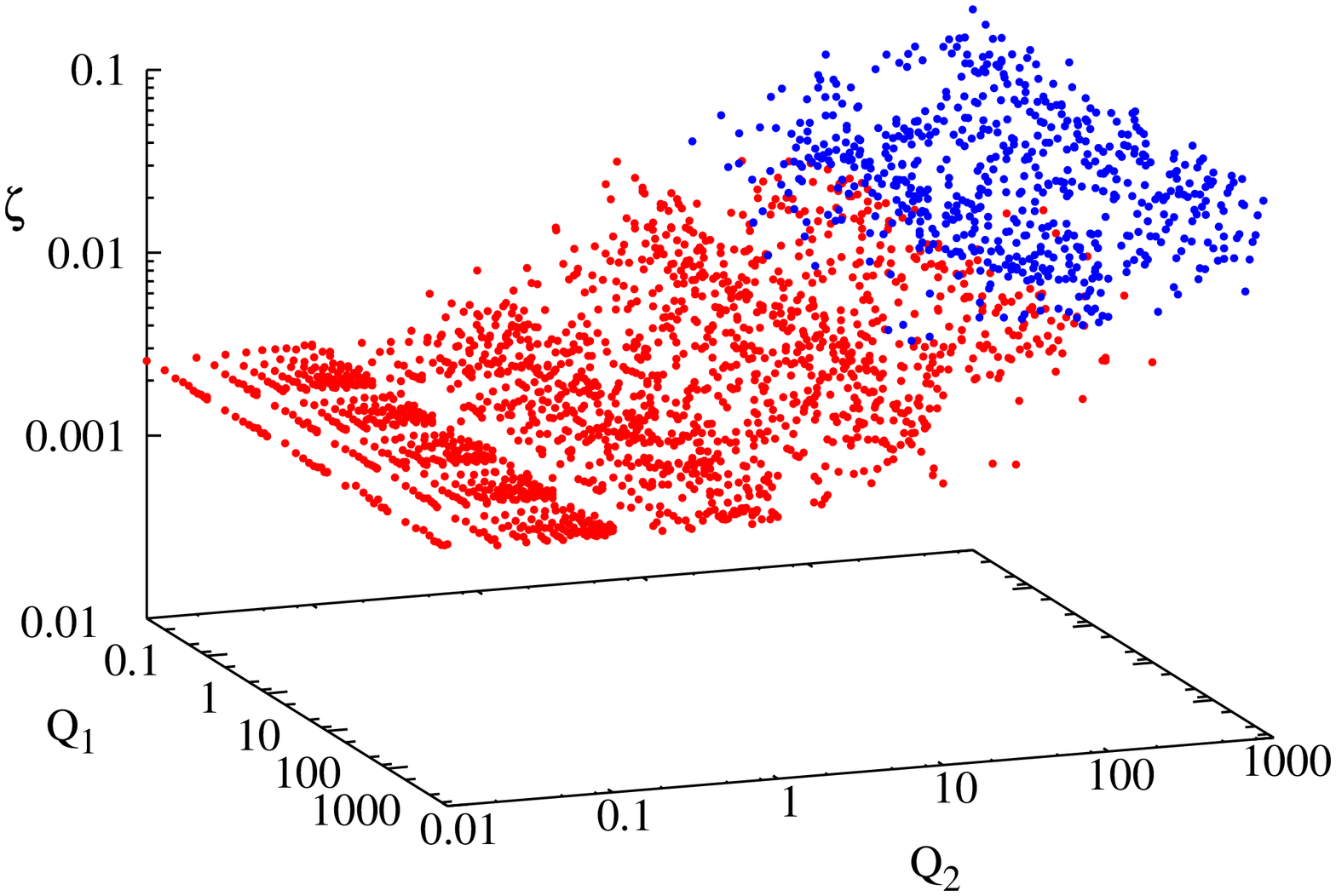}
\includegraphics[scale=0.3, angle = -90]{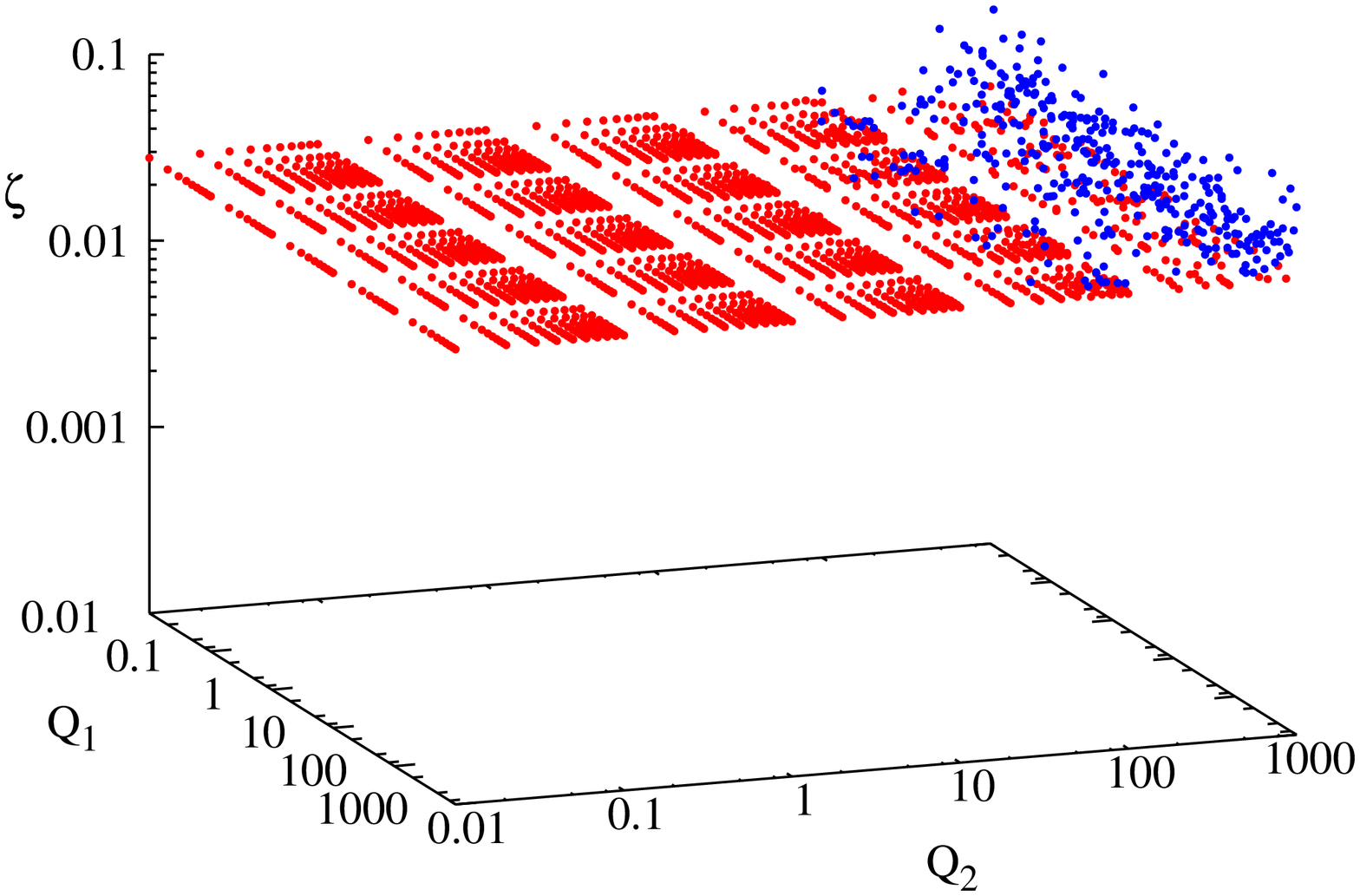} \\
\caption{$L^1$ error (see Eq.\ (\ref{eq:L1norm}) of the
position distribution $P(q)$ for the quartic potential
in Eq.\ (\ref{eq:quart_pot}) for different values of
the parameters $Q_1$, $Q_2$, $\gamma$, and the number
of RESPA steps. Top:  $\gamma = 0.1$ with
10 RESPA steps (left) and 100 RESPA steps (right).
Middle:  $\gamma = 1.0$ with 
10 RESPA steps (left) and 100 RESPA steps (right).
Bottom:  $\gamma = 10.0$ with 
10 RESPA steps (left) and 100 RESPA steps (right).
Red and blue dots separate values of $\zeta$ into
two regions:  For 10 RESPA steps, $\zeta < 0.02$
is designated with red dots, $\zeta > 0.02$ is
designated with blue dots.  For 100 RESPA steps,
the dividing value is $\zeta = 0.03$.
}
\label{fig:err_q1q2}
\end{figure}

A simple example illustrating the efficacy of the SIN(R) scheme 
is a harmonic oscillator of unit mass with an additional weak quartic perturbation,
for which the potential is
\begin{equation}
U(q) = {1 \over 2}\omega^2 q^2 + {1 \over 4}g q^4
\label{eq:quart_pot}
\end{equation}
The simple problem will serve to show that the most basic resonance
phenomenon has been eliminated in the SIN(R) scheme just as it was in the
original INR approach of 
Ref.~\onlinecite{Minary_PRL}.  For this problem, the large or outer
time step is set to the resonance time step $\pi/\omega$.  Figure~\ref{fig:quart_pot} (left)
shows the probability distribution $P(q)$ of the coordinate $q$ for this problem
when $\dti = \dt/100$ and $L=1$.  The two distribution in this figure
correspond to a standard Nos\'e-Hoover chain~\cite{Martyna92}
RESPA calculation and one using SIN(R).  Runs are of length $10^9\dt$
and use $\gamma = 1$, $\omega = 3$, $g=0.1$, $Q_1 = Q_2 = 1$, and $\beta = 1$
We also test the sensitivity of the algorithm to the choices
of the outer time step $\dt$ by fixing $\dti$ and
varying $\dt$ in increments of 0.05 between $0.01\pi/\omega$
and $1.2\pi/\omega$ and plotting the $L^1$ error
\begin{equation}
\zeta = {1 \over N_b}\sum_{i=1}^{N_b}\left|P(q_i)-P_{\rm an}(q_i)\right|
\label{eq:L1norm}
\end{equation}
in Fig.~\ref{fig:quart_pot}.  We see that the error is essentially 
monotonic over the entire range, suggesting that if any resonances 
exist (apart from the expected resonance at $\pi/\omega$) in this interval, the
method is not affected by them.  We also test the
sensitivity of the error to the values of
$Q_1$, $Q_2$, and $\gamma$ by plotting the $L^1$ error
for different values of these parameters in Fig.~\ref{fig:err_q1q2}.  
Here, $N_b$ is the number of bins used to generate the 
distribution, $q_i$ is the value of the coordinate in the $i$th bin, 
and $P_{\rm an}(q)$ is the analytical distribution $P_{\rm an}(q) \propto \exp(-\beta U(q))$.
It can be seen that for a wide range of parameter values, the
algorithm generates the same result with little change in the
error, suggesting that within a reasonable range of values, the
choice of parameters is not especially important.  

\subsection{A flexible water model}

\begin{figure}[htpb]
\includegraphics[scale=0.18]{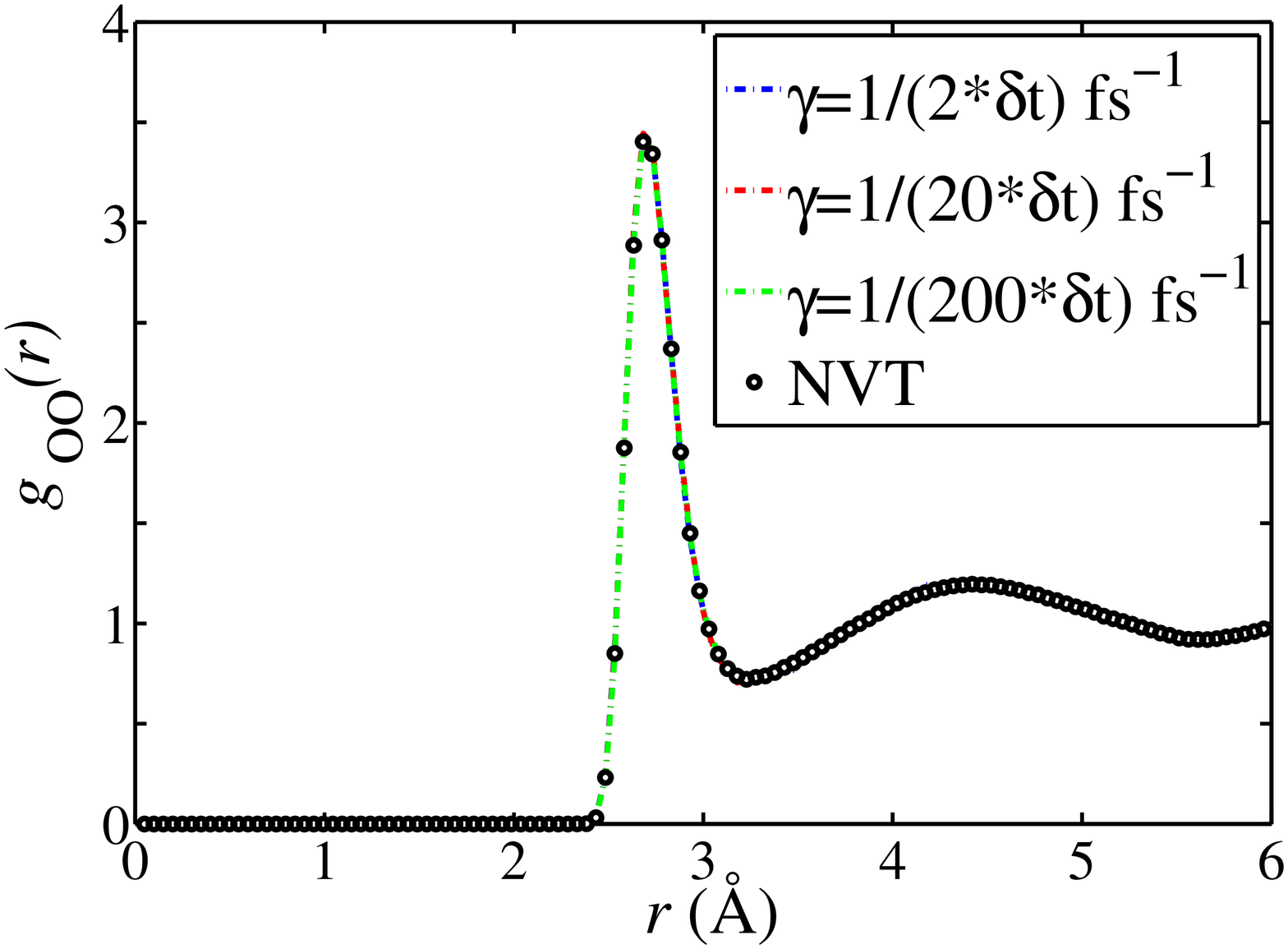}
\includegraphics[scale=0.18]{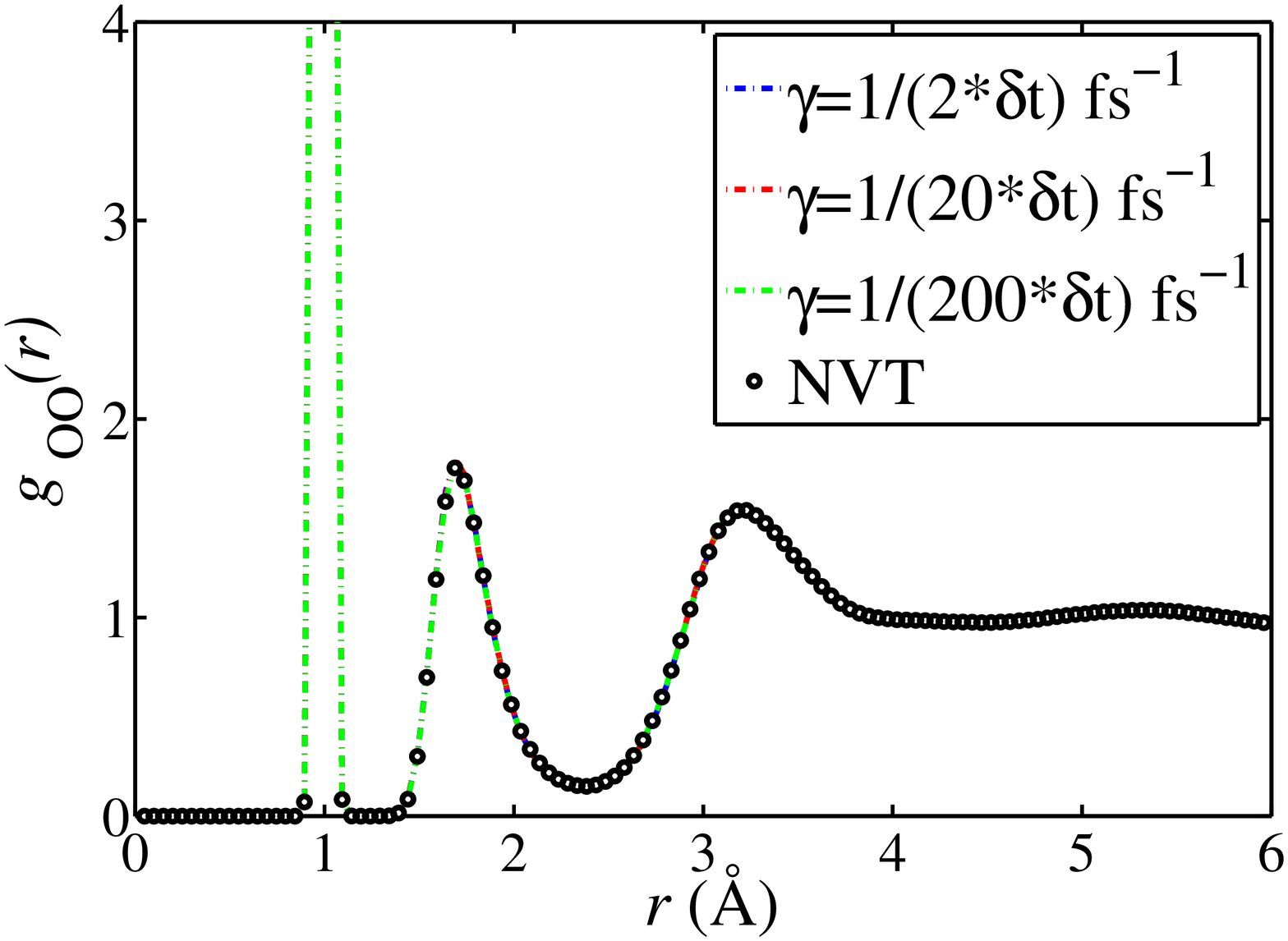}
\includegraphics[scale=0.18]{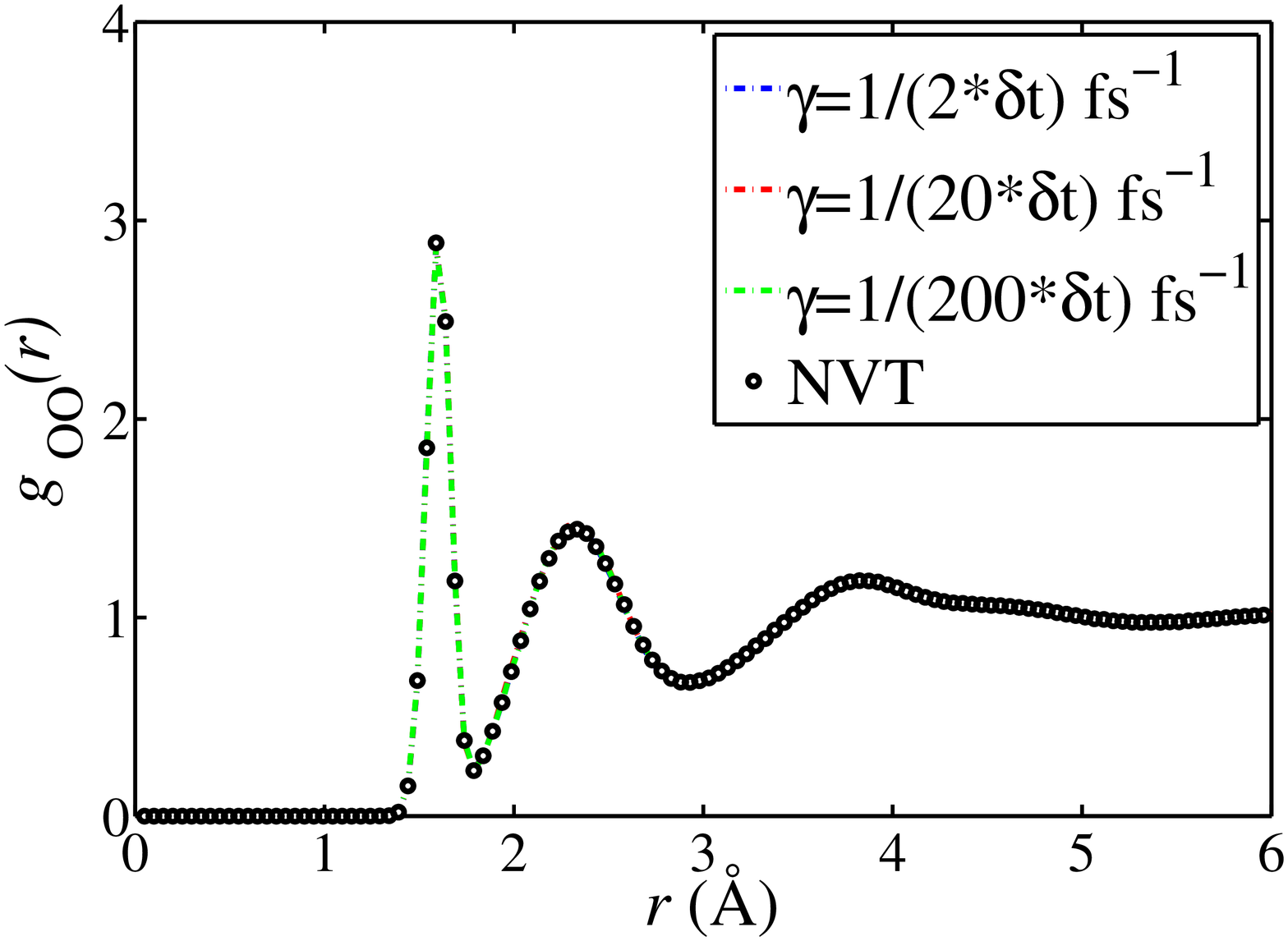}
\caption{Oxygen-oxygen (left), oxygen-hydrogen (middle) and
hydrogen-hydrogen (right) radial distribution functions for  $\gamma$
values of 0.5/$\delta t$, 0.05/$\delta t$, and 0.005/$\delta t$, where
$\delta t = 0.5$ fs.}
\label{fig:RDFs}
\end{figure}

\begin{figure}[htpb]
\includegraphics[scale=0.18]{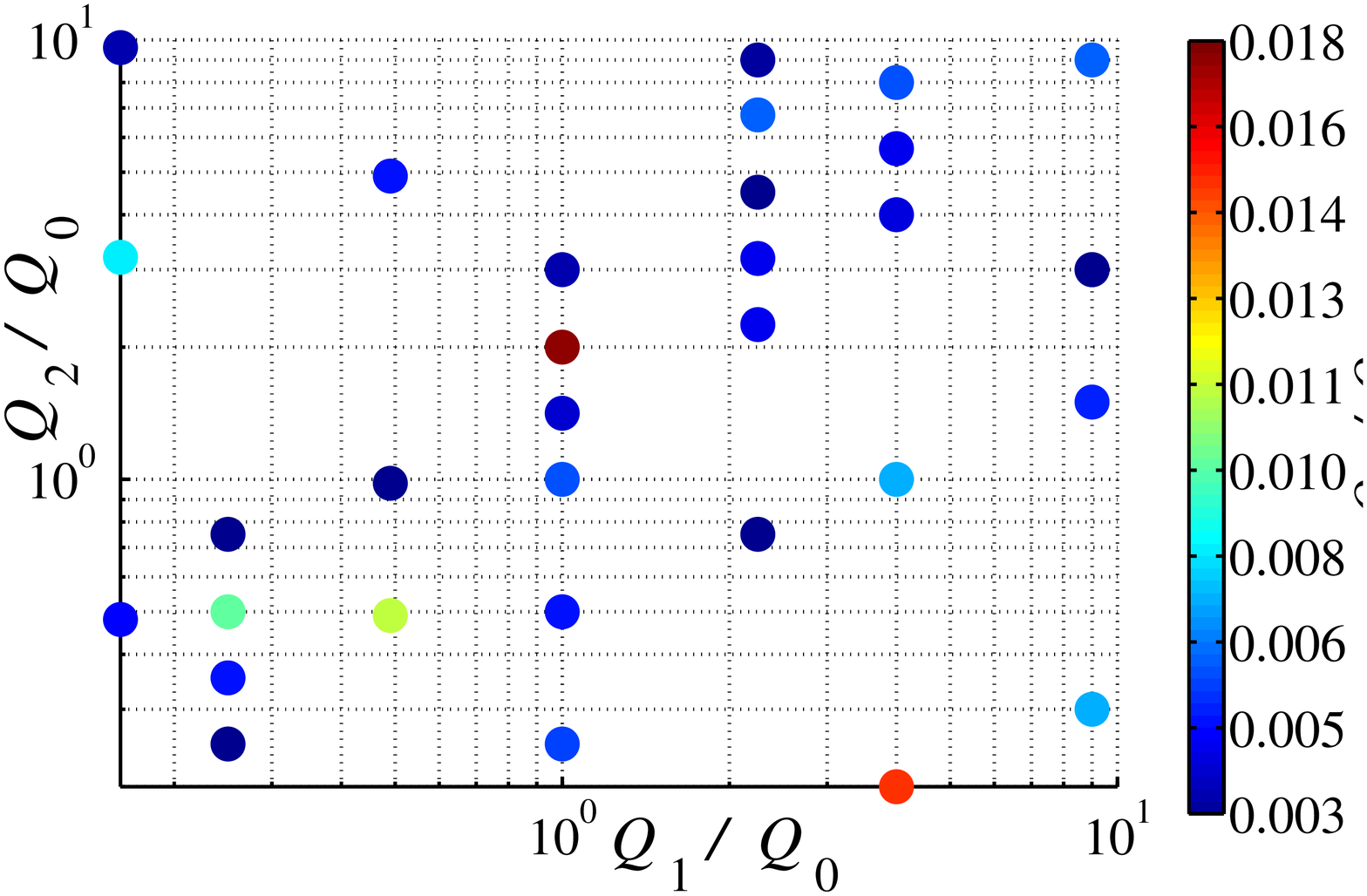}
\includegraphics[scale=0.18]{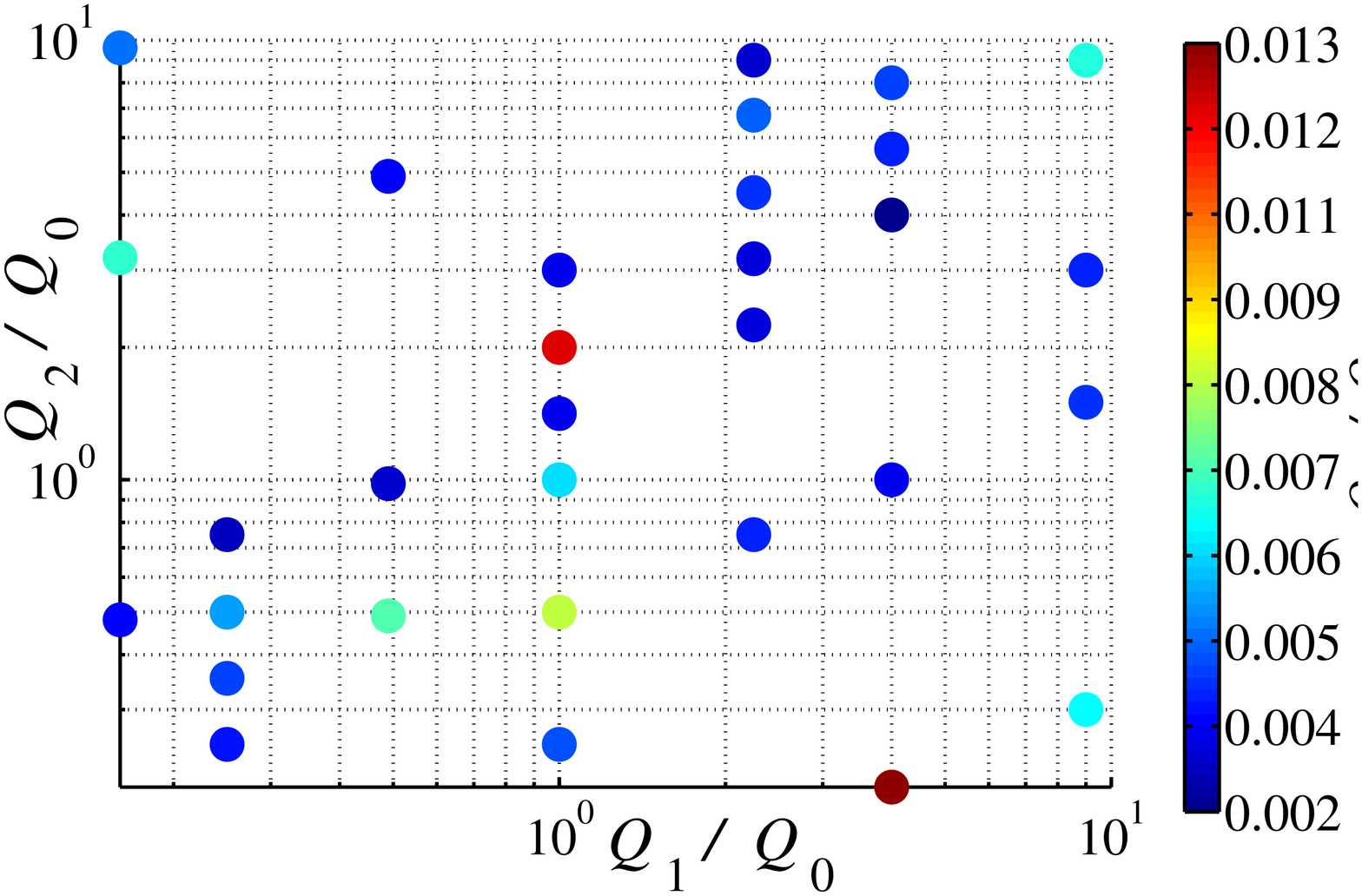}
\includegraphics[scale=0.18]{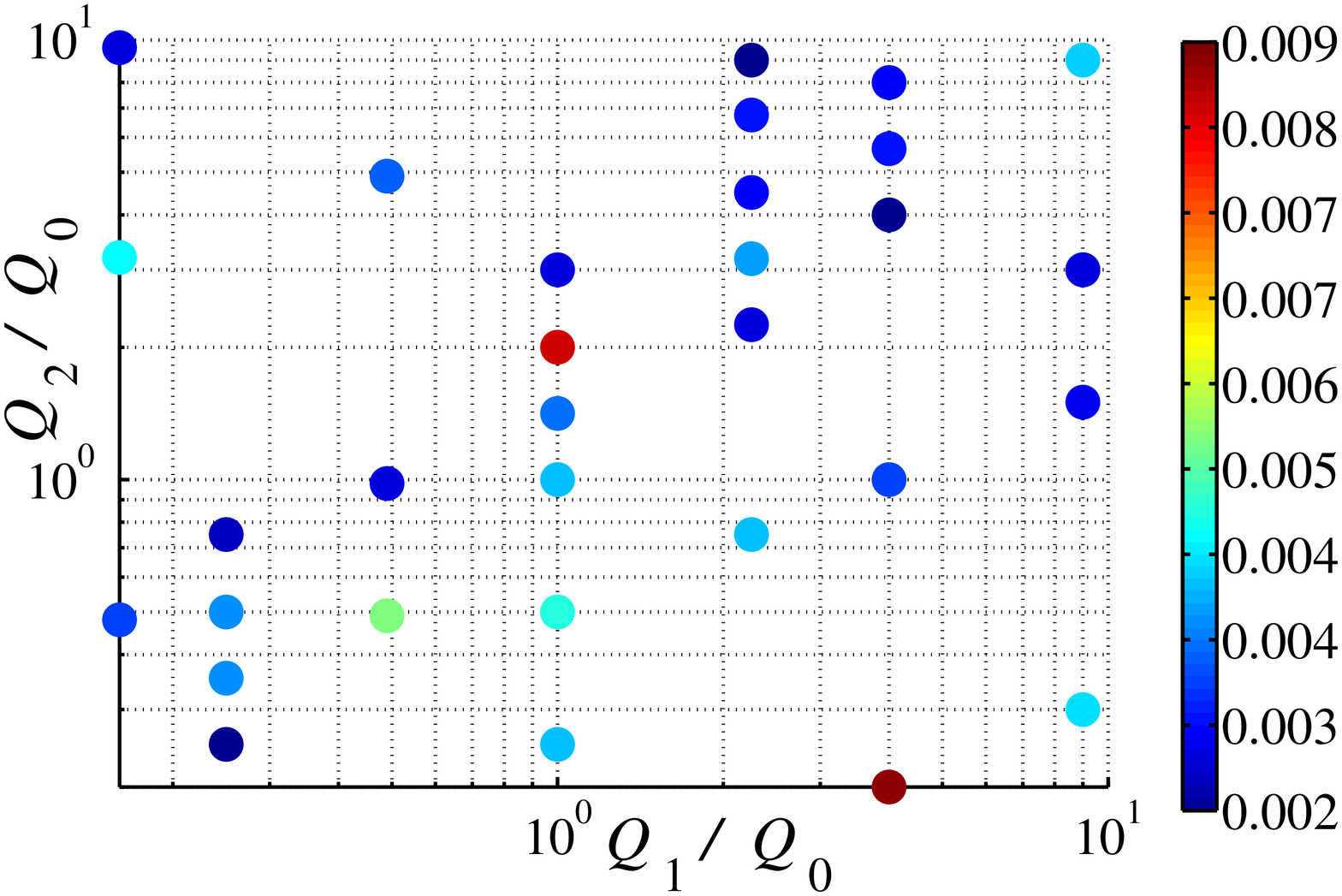}
\caption{$L^1$ error in the three RDFs for different values of
$Q_1$ and $Q_2$.  In these simulations, $\gamma = 0.05/\delta t$, 
where $\delta t = 0.5$ fs.}
\label{fig:errs}
\end{figure}

A challenging problem with a very wide separation of time 
scales is liquid water described by a fully flexible model~\cite{Paesani_06}.  
We first demonstrate the use of SIN(R) without RESPA in order to explore
the sensitivity of the method as a thermostatting approach to the
choice of parameters.  In Fig.~\ref{fig:RDFs}, we show the 
oxygen-oxygen, oxygen-hydrogen, and hydrogen-hydrogen radial distribution
functions for this model for different values of $\gamma$ in a system
of 512 molecules in a box of length 25 \AA\ subject to periodic 
boundary conditions.
In these simulations, the time step is 0.5 fs, and $Q_1 = Q_2 = k_{\rm B}T\tau^2$
where $T$ is the temperature of the simulation, i.e., 300 K,
$\tau = 10$ fs, and $L=4$.  Electrostatic interactions are treated using
the smooth particle-mesh Ewald method (SPME)~\cite{SPME}.  The SIN(R)
RDFs are shown compared to a benchmark set of RDFs generated
in the NVT ensemble using a Nos\'e-Hoover chain (NHC) thermostat~\cite{Martyna92}.
on each degree of freedom and a time step of 0.5 fs.  Simulation lengths range
between 300 and 600 ps and
are run using the PINY\_MD code~\cite{piny}.
It can be seen from the figure that the results are not sensitive to the
value of the friction parameter in the chosen range.  As a further test
of robust against the parameter choice, we show the $L^1$ error of 
the three radial distribution functions relative to NVT benchmark.
These are shown in Fig.~\ref{fig:errs}.
It can be see that for a wide range of values of $Q_1$ and $Q_2$ the
error is small and not particularly sensitive to specific values
of these parameters.  Not unexpectedly, however, as $Q_2$ increases,
the error does as well, since in this limit, the stochastic 
kicks have a smaller influence, and convergence becomes slower.

Because of resonance, MTS methods for flexible water have not been
able to achieve outer time steps larger than about 5 fs~\cite{Schlick_98,Ma_03}.
Here, we divide the forces into three time scales
corresponding to intramolecular motion (bond and bend
vibrations), short-range intermolecular interactions,
which include all non-bonded interactions within
a cutoff $r_{\rm sr}$, and long-range interactions.
For the latter, we consider two types:  First,
we consider the scheme suggested in Refs.~\onlinecite{Stuart96,Morrone_2011}, in which 
the long-range potential is expressed using Ewald summation
for the electrostatic interaction, and is given by
\begin{eqnarray}
U^{(2)}_{\rm long}(\ur) & = & U_{\rm recip}(\ur;\alpha,k_{\rm max})
\nonumber \\
\nonumber \\
& - & 
\sum_{{\bf S}}\sum_i \sum_{j\geq i}
\left(1 - \delta_{ij}^{(0)}\right)
\left(1 - \theta\left(r_{ij,{\bf S}} - r_{\rm cut}\right)\right)
{q_i q_j {{\rm erf}(\alpha r_{ij,{\bf S}}) \over r_{ij,{\bf S}}}}
\label{eq:long1}
\end{eqnarray}
where the sum over ${\bf S}$ is a sum over all periodic images,
$r_{ij,{\bf S}} = |\ur_i - \ur_j + {\bf S}|$, 
$\delta_{ij}^{(0)} = 1$ only if $i=j$ and ${\bf S} = (0,0,0)$,
$\theta(x)$ is the Heaviside step function, $q_i$ is the charge
on atom $i$, ${\rm erf}(x)$ is the error function,
$U_{\rm recip}(\ur)$ is the reciprocal-space contribution to the
Ewald sum for electrostatic interactions, the parameter
$\alpha$ determines the range of the real-space part
of the Ewald sum, $k_{\rm max}$ is the maximum magnitude
of the reciprocal-space lattice vectors used to evaluate
the long-range part of the electrostatic interaction.  We
refer to this subdivision as RESPA2.  The
second choice divides both the real and reciprocal space
sums into short and long-range contributions.  The long range
term then becomes
\begin{eqnarray}
U^{(1)}_{\rm long}(\ur) & = & U_{\rm recip}(\ur;\alpha,k_{\rm res})
\nonumber \\
\nonumber \\
& + & 
\sum_{{\bf S}}\sum_i \sum_{j\geq i}
\left(1 - \delta_{ij}^{(0)}\right)
\theta\left(r_{ij,{\bf S}} - r_{\rm cut}\right)
{q_i q_j {{\rm erf}(\alpha r_{ij,{\bf S}}) \over r_{ij,{\bf S}}}}
\label{eq:long2}
\end{eqnarray}
where $k_{\rm res} < k_{\rm max}$ is a reciprocal-space cutoff
that picks out reciprocal-space vectors with small magnitudes,
corresponding to the long-range contributions. We refer to this
subdivion as RESPA1.  Eq.\ (\ref{eq:long2})
might offer some advantages over Eq.\ (\ref{eq:long1}) in that it
does not rely on a potentially imperfect cancellation between real- and
reciprocal-sapce short-range contributions, which can 
cause numerical issues in Ewald summation in flexible systems~\cite{Martyna98}.

In the present simulations, SIN(R) is used with time steps
of $\delta t = $0.5 fs for the intramolecular interactions, 
$\Delta t = $ 3.0 fs for the short-range interactions, and the
outermost time step $\Delta T$ is allowed to vary 
in order to see how large it can be chosen without degrading
physical observables.  In all simulations, we choose the
values of the $Q_1$ and $Q_2$ using a value of $\tau = 10$ fs.

\begin{figure}[htpb]
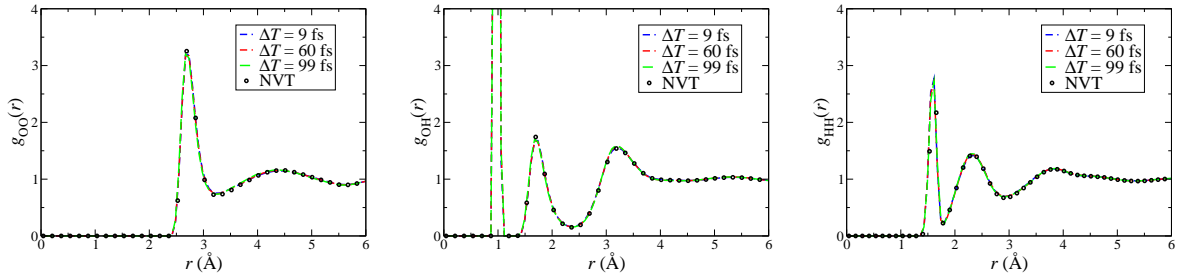

\vspace{0.2in}
\includegraphics[scale=0.2]{gOO_RESPA1.eps} \hspace*{0.1in}
\includegraphics[scale=0.2]{gOH_RESPA1.eps} \hspace*{0.1in}
\includegraphics[scale=0.2]{gHH_RESPA1.eps}
\caption{Oxygen-oxygen (left), oxygen-hydrogen (middle) and
hydrogen-hydrogen (right) radial distribution functions 
generated by SIN(R) for the RESPA1 scheme of 
Eq.\ (\ref{eq:long2}) with different choices for 
the outer time step $\Delta T$:  9 fs (blue), 
60 fs (red), 99 fs (green).  The benchmark NVT result
is shown as black circles.}
\label{fig:RDFs_R1}
\end{figure}

In Fig.~\ref{fig:RDFs_R1}, we show radial distribution functions (RDFs) 
corresponding to the RESPA1 scheme for outer time steps $\Delta T = $ 9 fs,
60 fs, and 99 fs using $L=4$, which allows us
to test robustness of SIN(R) with this parameter.  In addition,
we employ a friction $\gamma = 0.1$ fs$^{-1}$ and the
XI-RESPA scheme.  Dependence on $L$ and the choice
of XO-RESPA vs. XI-RESPA will be tested below.
In these calculations, the short-range cutoff is 6.0 \AA, and
the SPME reciprocal-space cutoff for the long-range interaction is half
that used for the full reciprocal-space cutoff.  The RDFs 
in Fig.~\ref{fig:RDFs_R1} are of comparable accuracy
to those in Ref.~\onlinecite{Minary_PRL} even at a time step
of 99 fs.  The SIN(R) RDFs are compared to those generated
from a benchmark simulation in the canonical ensemble using a 
single time step of 0.5 fs.  
\begin{figure}[htpb]
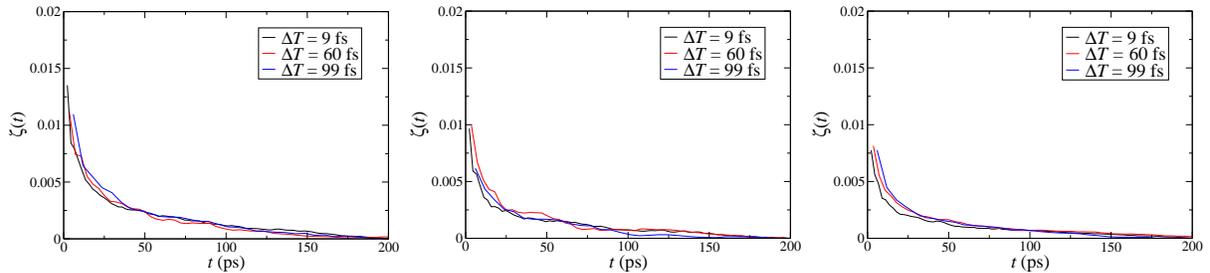

\vspace{0.2in}
\includegraphics[scale=0.2]{zeta_RESPA1_gOO.eps}
\includegraphics[scale=0.2]{zeta_RESPA1_gOH.eps}
\includegraphics[scale=0.2]{zeta_RESPA1_gHH.eps}
\caption{$L^1$ error $\zeta(t)$ relative to the fully 
converged value for each SIN(R) simulation shown in 
Fig.~\ref{fig:RDFs_R1}.  $\Delta T=$ 9 fs (black), 
$\Delta T = $60 fs (red), $\Delta T= $99 fs (blue). }
\label{fig:errs_R1}
\end{figure}
In Fig.~\ref{fig:errs_R1}, we plot the $L^1$ error
for each simulation relative to its final, converged value.  The
purpose of this comparison is to test whether the increase
in the outer time step $\Delta T$ and varying friction leads
to a change in the rate of convergence.  According to 
Fig.~\ref{fig:errs_R1}, the convergence rates of all
three simulations are similar, suggesting the different
choices of outer time step, friction, and value of $L$ do not strongly 
affect the convergence rate.  

\begin{figure}[htpb]
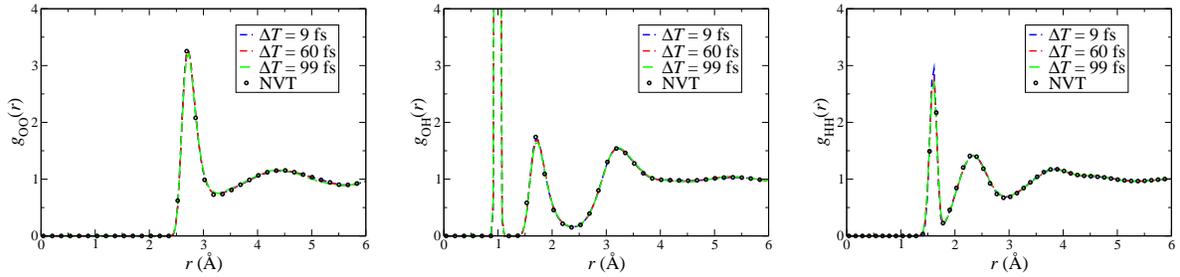

\vspace{0.2in}
\includegraphics[scale=0.2]{gOO_RESPA2.eps} \hspace*{0.1in}
\includegraphics[scale=0.2]{gOH_RESPA2.eps} \hspace*{0.1in}
\includegraphics[scale=0.2]{gHH_RESPA2.eps}
\caption{Oxygen-oxygen (left), oxygen-hydrogen (middle) and
hydrogen-hydrogen (right) radial distribution functions 
generated by SIN(R) for the RESPA2 scheme of 
Eq.\ (\ref{eq:long1}) with different choices for 
the outer time step $\Delta T$:  6 fs (blue), 
60 fs (red), 99 fs (green).  The benchmark NVT result
is shown as black circles. As in the RESPA1 example, all 
simulations employ a value of $L=4$.}
\label{fig:RDFs_R2}
\end{figure}

\begin{figure}[htpb]
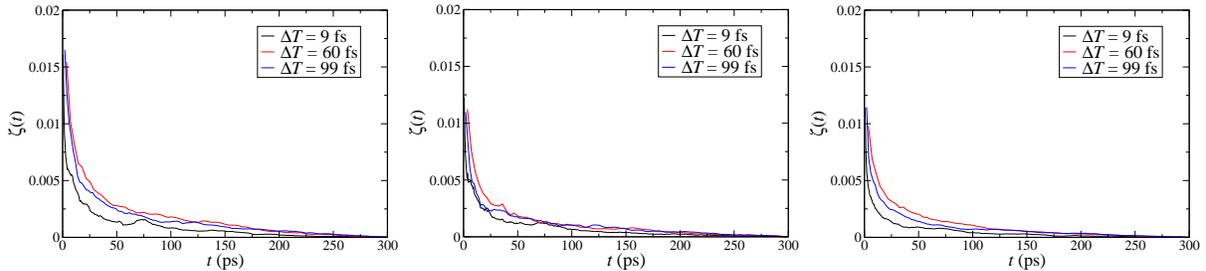

\vspace{0.2in}
\includegraphics[scale=0.2]{zeta_RESPA2_gOO.eps}
\includegraphics[scale=0.2]{zeta_RESPA2_gOH.eps}
\includegraphics[scale=0.2]{zeta_RESPA2_gHH.eps}
\caption{$L^1$ error $\zeta(t)$ relative to the fully 
converged value for each SIN(R) simulation shown in 
Fig.~\ref{fig:RDFs_R2}.  $\Delta T=$ 9 fs (black), 
$\Delta T = $60 fs (red), $\Delta T= $99 fs (blue). }
\label{fig:errs_R2}
\end{figure}

The RDFs for RESPA2 are shown in Fig.~\ref{fig:RDFs_R2},
here using outer time steps of $\Delta T = 6$ fs, 60 fs, and 99 fs.  
It is important to note that the RESPA2 scheme is
somewhat sensitive to the choice of the short-range
cutoff length, which we take to be 8 \AA\ in the 
present simulations.  The RESPA2 scheme also requires
a smooth switch to be applied to the forces~\cite{Morrone_2010},
and the stability of the scheme is sensitive to the
order of the switch.  Here, we employ a quintic switching function.
The corresponding $L^1$ error plots are shown in Fig.~\ref{fig:errs_R2}.
As these are somewhat more featured than the $L^1$ error plots of
Fig.~\ref{fig:errs_R1}, we extend the time axis somewhat
compared to that of Fig.~\ref{fig:errs_R1}.
Fig.~\ref{fig:errs_R2} again show that the rate of convergence is not 
strongly affected by the length of the outer time step, provided the
parameters are carefully chosen~\cite{Morrone_2010}.
Interestingly, in our implementation, the computational overhead
of RESPA1 with its shorter short-range cutoff and incorporation
of reciprocal space into the short-range reference system
is similar to that of RESPA2 with a purely real-space
short-range reference system and larger short-range cutoff.  
Clearly, however, implementation details will influence this balance,
and the use of massively parallel FFTs or GPUs for real-space
force calculations could reduce the overhead of the reference system,
thereby improving the efficiency of r-RESPA integration schemes.

\begin{figure}[htpb]
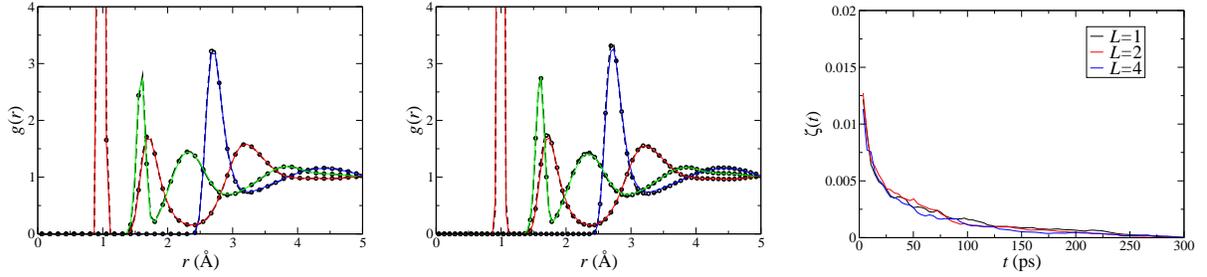

\vspace{0.2in}
\includegraphics[scale=0.2]{gr_RESPA1_L.eps} \hspace*{0.1in}
\includegraphics[scale=0.2]{gr_RESPA2_L.eps} \hspace*{0.1in}
\includegraphics[scale=0.2]{zeta_RESPA1_gOO_L.eps}
\caption{Complete set of radial distribution functions
for RESPA1 (left) and RESPA2 (middle) for
$L=1$, $L=2$, and $L=4$.  The $L=4$ RDFs are 
shown as blue (oxygen-oxygen), red (oxygen-hydrogen),
and green (hydrogen-hydrogen) solid lines.  
$L=1$ is shown with the dashed line, and
$L=2$ is shown as circular symbols.  In the right panel, we
show the $L^1$ error for the oxygen-oxygen RDF for the 
RESPA1 scheme.}
\label{fig:RDFs_L}
\end{figure}

We next examine the dependence on the choice of $L$ by showing 
the RDFs for RESPA1 and RESPA2 with $L=1$, $L=2$, 
and $L=4$ in Fig.~\ref{fig:RDFs_L}.
In these simulations $\Delta T = 60$ fs, and $\gamma = 0.1$ fs$^{-1}$.  The 
$L^1$ error plot in the right panel corresponds to the RESPA1 case and
illustrates that convergence is also insensitive to the choice of $L$.

\begin{figure}[htpb]
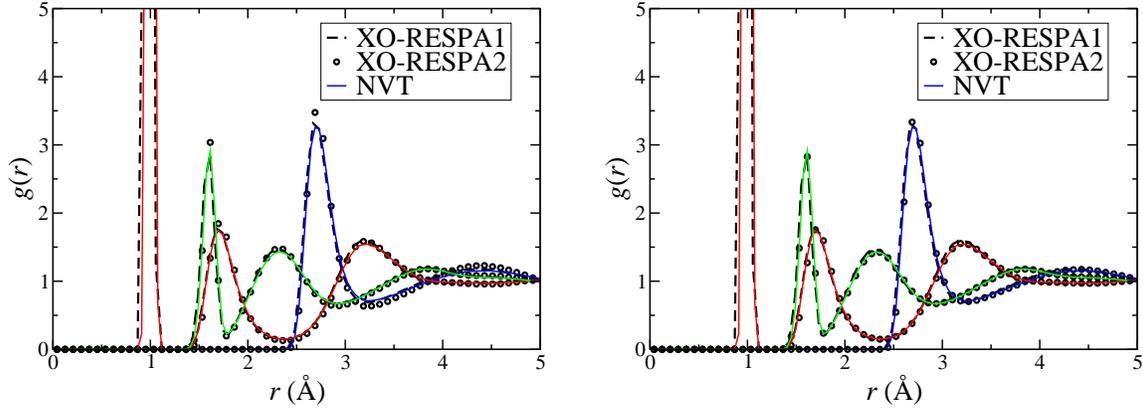

\vspace{0.2in}
\includegraphics[scale=0.3]{gr_XO_XI_60.eps} \hspace*{0.2in}
\includegraphics[scale=0.3]{gr_XO_XI_30.eps}
\caption{RDFs generated using XO-RESPA1 (dashed line) and XO-RESPA2 (symbols), compared to the NVT benchmark
(colored solid lines -- oxygen-oxygen(blue), oxygen-hydrogen (red), hydrogen-hydrogen (green). 
{\bf Left}:  $\Delta T = $ 60 fs for XO-RESPA2; {\bf Right}:  $\Delta T = $ 30 fs
for XO-RESPA2.}
\label{fig:XO_XI}
\end{figure}

Finally, we show that both XI-RESPA and XO-RESPA are capable of 
reproducing the RDFs.  This is illustrated in Fig.~\ref{fig:XO_XI},
which desplays RDFs generated XO-RESPA1 and XO-RESPA2 and comparing
these to the NVT benchmark results.
For these simulations, $L=4$, $\gamma = 0.1$ fs$^{-1}$, and $\Delta T = 60$ fs.
Here, we can see that, while both are in reasonable agreement with the NVT
benchmark, the RESPA2 results show a slightly larger deviation than the
corresponding XI-RESPA2 RDFs, illustrating that this scheme is somewhat more
sensitive to the choice of the simulation parameters, as alluded to previously.
We have also run this case with an outer time step of $\Delta T = 30$ fs and
have found that the deviations in Fig.~\ref{fig:XO_XI} largely disappear (see Fig.~\ref{fig:XO_XI}).

\section{Conclusions}
\label{sec:concl}

We have introduced a stochastic resonant-free multiple time step
approach for molecular dynamics calculations involving forces
that drive motion on many time scales.  The method builds on the
previous work of Minary {\it et al.}~\cite{Minary_PRL} and is shown
to allow very large time time steps to be employed for the slowest
forces, similar to what was obtained previously~\cite{Minary_PRL}.
The new stochastic version of the algorith, termed SIN(R), 
both simplifies the implementation of the method and can be shown
to be ergodic.  We presented the details of a multiple time step algorithm for the
equations of motion and demonstrated the performance on both
a simple, illustrative problem and a more realistic flexible
water model.  It was shown that time steps as large as 100 fs
could be employed for the long-range forces in a flexible water model.
The results display a slight sensitivity to the choice of force
subdivision, indicating that when very large time steps are used
some care must be exercised in choosing how short and long range
intermolecular forces are defined.  We have demonstrated
the performance of the approach for a typical fixed-charge model.
However, we also expect the present 
approach to be useful in simulation of complex systems
employing polarizable models and could even
enhance the efficiency of {\it ab initio} molecular dynamics
calculations. Investigations into the possible utility of the
present approach in calculations of this type will be the subject
of future research.

\acknowledgments
The authors gratefully acknolwedge useful discussions with Dr. Joseph A. Morrone.
M.E.T. and D. T. M. acknowledge support from NSF CHE-1012545 and CHE-1301314.
B.J.L acknowledges the support of the CWI (Amsterdam) and an NWO (Netherlands) ``visitors grant'' 
as well as useful conversations with A. Davie.

%
%
%
%
\clearpage
\bibliography{isok}
\bibliographystyle{jcp}

\end{document}